\documentclass[12pt]{article}
\usepackage{amsmath,amsfonts, amssymb}
\usepackage{graphicx}
\usepackage{longtable}
\usepackage{url}
\usepackage{tocloft}
\makeatletter
 
  \@addtoreset{equation}{section}
\makeatother
\def\beqar {\begin{eqnarray}}
\def\eeqar {\end{eqnarray}}
\def\beq {\begin{equation}}
\def\eeq {\end{equation}}
\def\A{{\cal A}}
\def\B{{\cal B}}
\def\C{{\cal C}}

\def\F{{\cal F}}

\def\H{{\cal H}}

\def\N{{\cal N}}

\def\S{{\cal S}}

\def\W{{\cal W}}

\def\al{\alpha}
\def\bt{\beta}
\def\del{\delta}
\def\Del{\Delta}
\def\ga{\gamma}
\def\Ga{\Gamma}
\def\ka{\kappa}
\def\ep{\epsilon}
\def\la{\lambda}

\def\om{\omega}
\def\Om{\Omega}
\def\th{\theta}
\def\et{\eta}
\def\si{\sigma}

\def\d{\partial}

\def\Ad{{\dot A}}
\def\Bd{{\dot B}}

\def\bu{{\bar u}}

\def\bth{{\bar \theta}}

\def\hf{\frac{1}{2}}

\def\<{\langle}\def\bra{\langle}
\def\>{\rangle}\def\ket{\rangle}

\def\Tr{{\rm Tr}}

\def\Path{{\rm P}}

\def\diag{{\rm diag}}

\def\det{{\rm det}}\def\dim{{\rm dim}}
\def\cp{{\bf CP}}

\begin{document}

\begin{titlepage}
\null\vspace{-62pt} \pagestyle{empty}
\begin{center}
\vspace{1.0truein}

{\Large\bf A note on generalized hypergeometric functions, \\
\vspace{.35cm}
KZ solutions, and gluon amplitudes} \\

\vspace{1.0in} {\sc Yasuhiro Abe} \\
\vskip .12in {\it Cereja Technology Co., Ltd.\\
3-11-15 UEDA-Bldg. 4F, Iidabashi   \\
Chiyoda-ku, Tokyo 102-0072, Japan }\\
\vskip .07in {\tt abe@cereja.co.jp}\\
\vspace{1.3in}

\centerline{\large\bf Abstract}
\end{center}
Some aspects of Aomoto's generalized hypergeometric functions
on Grassmannian spaces $Gr(k+1,n+1)$ are reviewed.
Particularly, their integral representations
in terms of twisted homology and cohomology are clarified
with an example of the $Gr(2,4)$ case which corresponds to
Gauss' hypergeometric functions.
The cases of $Gr(2, n+1)$ in general lead to $(n+1)$-point
solutions of the Knizhnik-Zamolodchikov (KZ) equation.
We further analyze the Schechtman-Varchenko integral
representations of the KZ solutions in relation to the
$Gr(k+1, n+1)$ cases.
We show that holonomy operators of the so-called KZ connections can
be interpreted as hypergeometric-type integrals.
This result leads to an improved description of
a recently proposed holonomy formalism for gluon amplitudes.
We also present a (co)homology interpretation of
Grassmannian formulations for scattering amplitudes
in ${\cal N} = 4$ super Yang-Mills theory.

\end{titlepage}
\pagestyle{plain} \setcounter{page}{2} 

\tableofcontents


\section{Introduction}

Recently, much attention is paid to Grassmannian formulations of
scattering amplitudes in $\N = 4$ super Yang-Mills theory.
The Grassmannian formulations are initially proposed (see, {\it e.g.},
\cite{oai:arXiv.org:0907.5418}-\cite{oai:arXiv.org:1006.1899})
to make it manifest that the $\N = 4$ super Yang-Mills amplitudes
are invariant under the dual superconformal symmetry \cite{Drummond:2008vq} at tree level.
For more recent developments, see, {\it e.g.},
\cite{Cachazo:2012pz}-\cite{Ferro:2015grk}.

As discussed in \cite{Ferro:2014gca,Ferro:2015grk}, these Grassmannian formulations
have revived interests in a purely mathematical subject, {\it i.e.},
generalized hypergeometric functions on Grassmannian spaces $Gr(k+1,n+1)$, which were
introduced and developed by Gelfand \cite{Gelfand:1986} and independently
by Aomoto \cite{Aomoto:1994bk} many years ago.
In relation to physics, it has been known that solutions of
the Knizhnik-Zamolodchikov (KZ) equation in conformal field theory
are expressed in terms of the generalized hypergeometric functions
\cite{Date:1989tg}-\cite{Awata:1991az}.
One of the main goals of this note is to present a clear and systematic
review on these particular topics in mathematical physics.
Particularly, we revisit integral representations of
the KZ solutions by Schechtman and Varchenko \cite{Schechtman:1989,Schechtman:1990zc}
and analyze them in terms of a bilinear construction of
hypergeometric integrals, using twisted homology and cohomology.
Along the way, we also consider in detail Gauss' original hypergeometric
functions in Aomoto's framework so as to familiarize ourselves to
the concept of twisted homology and cohomology.

Another goal of this note is to study and understand
analytic aspects of the holonomy operator of the so-called KZ connection.
The holonomy of the KZ connection is first introduced by Kohno
\cite{Kohno:2002bk} (see also Appendix 4 in \cite{Aomoto:1994bk})
as a monodromy representation of the KZ equation in a form
of the iterated integral \cite{Chen:1977oja}.
Inspired by Kohno's result and
and Nair's observation \cite{Nair:1988bq} on the maximally helicity violating (MHV)
amplitudes of gluons (also called the Parke-Taylor amplitudes \cite{Parke:1986gb})
in supertwistor space, the author has recently proposed a novel framework
of deriving gluon amplitudes \cite{Abe:2009kn}
where an S-matrix functional for the gluon amplitudes is defined
in terms of the holonomy operator of a certain KZ connection.
This framework, what we call the holonomy formalism, is
intimately related to braid groups and Yangian symmetries.
As mentioned in \cite{Abe:2011af}, the holonomy formalism
also suggests a natural origin of the dual conformal symmetries.
Towards the end of this note we would provide more rigorous mathematical
foundations of the holonomy formalism and present an improved description of it.
Lastly, we also consider the more familiar Grassmannian formulations of gluon amplitudes
in the same framework. Namely, we analyze integral representations
of the Grassmannian formulations and present a (co)homology interpretation
of those integrals.

This note is organized as follows. In the next section we review
some formal results of Aomoto's generalized hypergeometric functions on $Gr(k+1,n+1)$,
based on textbooks by Japanese mathematicians
\cite{Aomoto:1994bk,Yoshida:1997bk,Haraoka:2002bk,Kimura:2007bk}.
We present a review in a pedagogical fashion since these results are
not familiar enough to many physicists.
In section 3 we consider a particular case $Gr (2, n+1)$
and present its general formulation.
In section 4 we further study the case of $Gr(2,4)$
which reduces to Gauss' hypergeometric function.
Imposing permutation invariance among branch points, we here
obtain new realizations of the hypergeometric differential equation
in a form of a first order Fuchsian differential equation.

In section 5 we apply Aomoto's results to the KZ equation.
We first focus on four-point KZ solutions
and obtain them in a form of the hypergeometric integral.
We then show that $(n+1)$-point KZ solutions in general
can be represented by generalized hypergeometric functions on $Gr( 2, n+1)$.
We further consider the Schechtman-Varchenko
integral representations of the KZ solutions in this context.
The $(n+1)$-point KZ solutions can also be
represented by the hypergeometric-type integrals on $Gr( k+1, n+1)$
but we find that there exist ambiguities in the construction
of such integrals for $k \ge 2$.
In section 6 we review the construction of the holonomy operators
of the KZ connections. We make a (co)homology interpretation
of the holonomy operator and obtain a better understanding
of analytic properties of the holonomy operator.

The holonomy operator gives a monodromy representation of the
KZ equation, which turns out to be a linear representation of a braid group.
This mathematical fact has been one of the essential ingredients in the holonomy formalism for
gluon amplitudes. In section 7 we briefly review this holonomy formalism
and present an improved description of it. In section 8 we also consider
the Grassmannian formulations of gluon amplitudes. We observe that
these formulations can also be interpreted in terms of the hypergeometric-type integrals.
Lastly, we present a brief conclusion.

\section{Aomoto's generalized hypergeometric functions}
\noindent
\underline{Definition}

Let $Z$ be a $(k+1) \times (n+1)$ matrix
\beq
    Z = \left(
          \begin{array}{ccccc}
            z_{00} & z_{01} & z_{02} & \cdots & z_{0n} \\
            z_{10} & z_{11} & z_{12} & \cdots & z_{1n} \\
            \vdots & \vdots & \vdots &  & \vdots \\
            z_{k0} & z_{k1} & z_{k2} & \cdots & z_{kn} \\
          \end{array}
        \right)
    \label{1-1}
\eeq
where $k < n$ and the matrix elements are complex, $z_{ij} \in {\bf C}$
($0 \le i \le k \, ;~  0 \le j \le n $).
A function of $Z$, which we denote $F(Z)$, is defined as {\it
a generalized hypergeometric function on Grassmannian space $Gr (k+1, n+1)$
}
when it satisfies the following relations:
\beqar
    \sum_{j = 0}^{n} z_{ij} \frac{\d F}{\d z_{pj}} &=& - \del_{ip} F
    ~~~ ( 0 \le i, p \le k )
    \label{1-2a} \\
    \sum_{i = 0}^{k} z_{ij} \frac{\d F}{\d z_{ij}} &=&  \al_{j} F
    ~~~ ( 0 \le j \le n )
    \label{1-2b} \\
    \frac{\d^2 F}{\d z_{ip}\d z_{jq}} &=& \frac{\d^2 F}{\d z_{iq}\d z_{jp}}
    ~~~ ( 0 \le i, j \le k \,  ; ~ 0 \le p, q \le n  )
    \label{1-2c}
\eeqar
where the parameters $\al_j$ obey the non-integer conditions
\beqar
    \al_j & \not\in & {\bf Z} ~~~~~ ( 0 \le j \le n )
    \label{1-3a} \\
    \sum_{j = 0}^{n} \al_j & =& - (k + 1)
    \label{1-3b}
\eeqar

\vskip 0.5cm \noindent
\underline{Integral representation of $F(Z)$ and twisted cohomology}

The essence of Aomoto's generalized hypergeometric function
\cite{Aomoto:1994bk} is that, by use of the so-called
twisted de Rham cohomology,\footnote{
The {\it twisted} de Rham cohomology is a version of
the ordinary de Rham cohomology into which multivalued functions,
such as $\Phi$ in (\ref{1-5a}), are incorporated.
For mathematical rigor on this, see Section 2 in \cite{Aomoto:1994bk}.
}
$F(Z)$ can be written in a form of integral:
\beq
    F ( Z ) \, = \, \int_\Delta \Phi \om
    \label{1-4}
\eeq
where
\beqar
    \Phi &=& \prod_{j = 0}^{n} l_j (\tau )^{\al_j}
    \label{1-5a} \\
    l_j (\tau ) &=& \tau_0 z_{0j} + \tau_1 z_{1j}
    + \cdots + \tau_k z_{k j} ~~~~ ( 0 \le j \le n)
    \label{1-5b} \\
    \om &=& \sum_{i = 0}^{k} ( -1 )^i \tau_i
    d \tau_0 \wedge d \tau_1 \wedge \cdots \wedge
    d \tau_{i-1} \wedge d \tau_{i + 1} \wedge \cdots \wedge d \tau_k
    \label{1-5c}
\eeqar
The complex variables $\tau = ( \tau_0 , \tau_1 , \cdots , \tau_k )$
are homogeneous coordinates of the complex projective space $\cp^k$,
{\it i.e.}, ${\bf C}^{k+1 } - \{ 0, 0, \cdots , 0 \}$.
The multivalued function $\Phi$ is then defined in a space
\beq
    X \, = \, \cp^{k} - \bigcup_{j = 0}^{n} {\cal H}_j
    \label{1-6}
\eeq
where
\beq
    {\cal H}_j \, = \, \{  \tau \in \cp^k  \, ; ~ l_j (\tau ) = 0 \}
    \label{1-7}
\eeq

We now consider the meaning of the integral path $\Delta$.
Since the integrand $\Phi \om$ is a multivalued
$k$-form, simple choice of $\Delta$ as
a $k$-chain on $X$ is not enough.
{\it Upon the choice of $\Delta$, we need to implicitly
specify branches of $\Phi$ on $\Delta$ as well, otherwise
we can not properly define the integral.}
In what follows we assume these implicit conditions.

Before considering further properties of $\Delta$,
we here notice that $\om$ has an ambiguity
in the evaluation of the integral (\ref{1-4}).
Suppose $\al$ is an arbitrary $(k -1)$-form defined in $X$.
Then an integral over the exact $k$-form $d( \Phi \al )$ vanishes:
\beq
    0 \, = \, \int_\Delta d ( \Phi \al ) \, = \,
    \int_\Delta \Phi \left(
    d \al + \frac{d \Phi}{\Phi} \wedge \al
    \right) \, = \, \int_\Delta \Phi \nabla \al
    \label{1-8}
\eeq
where $\nabla$ can be interpreted as a covariant (exterior) derivative
\beq
    \nabla  \, = \, d  + d \log \Phi \wedge \, = \,
    d + \sum_{j = 0}^{n} \al_j \frac{ d l_j}{l_j} \wedge
    \label{1-9}
\eeq
This means that $\om^\prime = \om + \nabla \al$
is equivalent to $\om$ in the definition of the integral (\ref{1-4}).
Namely, $\om$ and $\om^\prime$ form an equivalent class,
$\om \sim \om^\prime$. This equivalent class is called the
cohomology class.

To study this cohomology class, we consider the differential equation
\beq
    \nabla f \, = \, d f + \sum_{j = 0}^{n} \al_j \frac{d l_j}{l_j} f \, =\, 0
    \label{1-10}
\eeq
General solutions are locally determined by
\beq
    f \, = \, \la \, \prod_{j = 0}^{n} l_{j} (\tau )^{- \al_j} ~~~~~~
    ( \la \in {\bf C}^{\times} )
    \label{1-11}
\eeq
These local solutions are thus basically given by $1 / \Phi$.
The idea of locality is essential since even if $1 / \Phi$ is multivalued
within a local patch it can be treated as a single-valued function.
Analytic continuation of these solutions forms a fundamental homotopy group
of a closed path in $X$ (or $1/X$ to be precise but it can
be regarded as $X$ by flipping the non-integer powers $\al_j$ in (\ref{1-5a})).
The representation of this fundamental group is called the {\it monodromy}
representation.
The monodromy representation determines the local system
of the differential equation (\ref{1-10}).
The general solution $f$ or $1 / \Phi$ gives a rank-1 local system in
this sense\footnote{
It is `rank-1' because each factor $l_j ( \tau )$ in the local solutions
(\ref{1-11}) is first order in the elements of $\tau$.
}.
We denote this rank-1 local system by ${\cal L}$.
The above cohomology class is then defined as an element
of the $k$-th cohomology group of $X$ over ${\cal L}$,
{\it i.e.},
\beq
    [ \om ] \in H^k ( X , {\cal L} )
    \label{1-12}
\eeq
This cohomology group $H^k ( X , {\cal L} )$ is also
called {\it twisted} cohomology group.

\vskip 0.5cm \noindent
\underline{Twisted homology and twisted cycles}

Having defined the cohomology group $H^k ( X , {\cal L} )$, we
can now define the dual of it, {\it i.e.}, the $k$-th
homology group $H_k ( X , {\cal L}^{\vee})$, known as the twisted
homology group, where ${\cal L}^{\vee}$ is the rank-1 dual local system
given by $\Phi$.
A differential equation corresponding to ${\cal L}^{\vee}$
can be written as
\beq
    \nabla^{\vee} g \, = \, d g -  \sum_{j = 0}^{n} \al_j \frac{d l_j}{l_j}
    g \, =\, 0
    \label{1-13}
\eeq
We can easily check that the general solutions are given by $\Phi$:
\beq
    g \, = \, \la \, \prod_{j = 0}^{n} l_{j} (\tau )^{ \al_j}
    \, = \, \la \Phi ~~~~~~
    ( \la \in {\bf C}^{\times} )
    \label{1-14}
\eeq
As before, an element of $H_k ( X , {\cal L}^{\vee})$
gives an equivalent class called a homology class.

In the following, we show that the integral path $\Del$
forms an equivalent class and see that it coincides with
the above homology class.
Applying Stokes' theorem to  (\ref{1-8}), we find
\beq
    0  \, = \, \int_\Delta \Phi \nabla \al \, = \, \int_{\d \Delta} \Phi \al
    \label{1-15}
\eeq
where $\al$ is an arbitrary $(k -1 )$-form as before.
The boundary operator $\d$ is in principle
determined from $\Phi$ (with information on branches).
Denoting $C_p ( X , {\cal L}^{\vee})$
a $p$-dimensional chain group on $X$ over ${\cal L}^{\vee}$,
we can express the boundary operator as
$\d \, : \, C_p ( X , {\cal L}^{\vee})  \longrightarrow
C_{p-1} ( X , {\cal L}^{\vee})$.
Since the relation (\ref{1-15}) holds for an arbitrary $\al$,
we find that the $k$-chain $\Del$ vanishes by the action of $\d$:
\beq
    \d \Delta \, = \, 0
    \label{1-16}
\eeq
The $k$-chain $\Del$ satisfying above is generically called the $k$-cycle.
In the current framework it is also  called the {\it twisted cycle}.
Since the boundary operator satisfies $\d^2 = 0$, the $k$-cycle
has a redundancy in it. Namely,
$\Del^\prime = \Del + \d C_{(+1)}$
also becomes the $k$-cycle where $C_{(+1)}$ is
an arbitrary $( k + 1 )$-chain or an element of $C_{k+1} ( X , {\cal L}^{\vee})$.
Thus $\Del$ and $\Del^\prime$ form an equivalent class,
$\Del \sim \Del^\prime$, and this is exactly the
homology class defined by $H_k ( X , {\cal L}^{\vee} )$, {\it i.e.},
\beq
    [ \Delta ] \in H_k ( X , {\cal L}^{\vee} )
    \label{1-17}
\eeq

To summarize, the generalized hypergeometric
function (\ref{1-4}) is determined by the following bilinear form
\beqar
    H_k ( X , {\cal L}^{\vee} ) \times H^k ( X , {\cal L} )
    & \longrightarrow & {\bf C}
    \label{1-18} \\
    \left( [ \Delta ], [ \om ] \right) & \longrightarrow &
    \int_{\Del} \Phi \om
    \label{1-19}
\eeqar

\vskip 0.5cm \noindent
\underline{Differential equations of $F(Z)$}

The condition $l_j (\tau ) = 0$ in (\ref{1-7}) defines
a hyperplane in $(k+1)$-dimensional spaces.
To avoid redundancy in configuration of hyperplanes,
we assume the set of hyperplanes are non-degenerate,
that is, we consider the hyperplanes in {\it general position}.
This can be realized by demanding that any $(k+1)$-dimensional
minor determinants of the $(k+1) \times (n+1)$ matrix $Z$ are nonzero.
We then redefine $X$ in (\ref{1-6}) as
\beq
    X =  \{
    Z \in Mat_{k+1 , n+1} ( {\bf C} ) | \mbox{ {\rm any $(k+1)$-dim
    minor determinants of $Z$ are nonzero}}
    \}
    \label{1-20}
\eeq
In what follows we implicitly demand this condition in $Z$.
The configuration of $n+1$ hyperplanes in $\cp^k$ is determined by
this matrix $Z$.

Apart from the concept of hyperplanes, we can also interpret that
the above $Z$ provides $n+1$ {\it distinct points} in $\cp^k$.
Since a homogeneous coordinate of $\cp^k$ is given by
${\bf C}^{k+1} - \{0, 0, \cdots, 0 \}$, we can consider each of
the $n+1$ column vectors of $Z$ as a point in $\cp^k$;
the $j$-th column representing the $j$-th homogeneous coordinates
of $\cp^k$ ($j = 0,1, \cdots , n$).

The scale transformation, under which
the $\cp^k$ homogeneous coordinates are invariant,
is realized by an action of $H_{n+1} = \{ \diag (h_0 , h_1 , \cdots h_n )
| h_j \in {\bf C}^\times \}$ from right on $Z$.
The general linear transformation of the
homogeneous coordinates, on the other hand,
can be realized by an action of $GL(k+1, {\bf C})$ from left.
These transformations are then given by
\beqar
    {\mbox {\rm Linear transformation:}}&~& Z \rightarrow Z^\prime = g Z
    \label{1-21a} \\
    {\mbox {\rm Scale transformation:}}&~& Z \rightarrow Z^\prime = Z h
    \label{1-21b}
\eeqar
where $g \in GL(k+1, {\bf C})$ and $h \in H_{n+1}$.
Under these transformations the integral $F(Z)$ in (\ref{1-4}) behaves as
\beqar
    F (g Z ) &=& (\det g)^{-1} F(Z)
    \label{1-22a}\\
    F ( Z h ) &=& F (Z) \prod_{j = 0}^{n} h_{j}^{\al_j}
    \label{1-22b}
\eeqar

We now briefly show that the above relations lead to the
defining equations of the generalized hypergeometric functions
in (\ref{1-2a}) and (\ref{1-2b}), respectively.
Let ${\bf 1}_n$ be the $n$-dimensional identity matrix ${\bf 1}_n
= \diag ( 1,1, \cdots ,1)$, and $E_{ij}^{(n)}$ be an
$n \times n$ matrix in which
only the $(i,j)$-element is 1 and the others are zero.
We consider $g$ in a particular form of
\beq
    g = {\bf 1}_{k+1} + \ep E^{(k+1)}_{pi}
    \label{1-23}
\eeq
where $\ep$ is a parameter.
Then $gZ$ remains the same as $Z$ except the $p$-th row
which is replaced by $(z_{p0} + \ep z_{i0}, z_{p1} + \ep z_{i1}, \cdots,
z_{pn} + \ep z_{in})$.
Then the derivative of $F(gZ)$ with respect to $\ep$ is expressed as
\beq
    \frac{\d}{\d \ep} F (gZ) \, = \, \sum_{j=0}^{n} z_{ij}
    \frac{\d }{\d z_{pj}} F(gZ)
    \label{1-24}
\eeq
On the other hand, using
\beq
    \det g = \left\{
    \begin{tabular}{l}
      $1 ~~~ (i \ne p)$\\
      $\ep ~~~ (i = p )$ \\
    \end{tabular}
    \right.
    \label{1-25}
\eeq
and (\ref{1-22a}), we find
\beq
    \frac{\d}{\d \ep} F (gZ) =
\left\{
    \begin{tabular}{l}
      $0 ~~~~~~~~~~~~~ (i \ne p)$\\
      $- \frac{1}{\ep^2} F(Z) ~~~ (i = p )$ \\
    \end{tabular}
    \right.
    \label{1-26}
\eeq
Evaluating the derivative at $\ep = 0$ and $\ep = 1$ for
$i \ne p$ and $i = p$, respectively, we then indeed find that
(\ref{1-22a}) leads to the differential equation (\ref{1-2a}).

Similarly, parametrizing $h$ as
\beq
    h = \diag( h_0 , \cdots , h_{j-1} , (1+ \ep )h_j  , h_{j+1} , \cdots , h_n )
    \label{1-27}
\eeq
with $0 \le j \le n$, we find that $Zh$ has only one $\ep$-dependent
column corresponding to the $j$-th column, $\left( z_{0j}( 1 +\ep )h_j  ,
z_{1j} ( 1 +\ep )h_j, \cdots , z_{kj} ( 1 +\ep )h_j \right)^{T}$.
The derivative of $F(Zh)$ with respect to $\ep$ is then expressed as
\beq
    \frac{\d}{\d \ep} F (Zh) \, = \, \sum_{i=0}^{k} z_{ij}
    \frac{\d }{\d z_{ij}} F(Zh) \, = \, \sum_{i=0}^{k} z_{ij}
    \frac{\d }{\d z_{ij}} F(Z)(1+ \ep )^{\al_j} \prod_{ l = 0}^{n} h_{l}^{\al_l}
    \label{1-28}
\eeq
where in the last step we use the relation from (\ref{1-22b}):
\beq
    F (Zh ) \, = \, F(Z) (1+ \ep )^{\al_j} \prod_{ l = 0}^{n} h_{l}^{\al_l}
    \label{1-29}
\eeq
The same derivative can then be expressed as
\beq
    \frac{\d}{\d \ep} F (Zh) \, = \, \al_j F(Z) (1+ \ep )^{\al_j  - 1}
    \prod_{ l = 0}^{k} h_{l}^{\al_l}
    \label{1-30}
\eeq
Setting $\ep = 0$, we can therefore derive the equation (\ref{1-2b}).

The other equation (\ref{1-2c}) for $F(Z)$ follows from the
definition of $\Phi$. From (\ref{1-5a}) and (\ref{1-5b}) we find
that $\Phi$ satisfies
\beq
    \frac{\d \Phi}{\d z_{ip} } \, = \, \frac{ \al_i \tau_p}{l_i (\tau) } \Phi
    \label{1-31}
\eeq
This relation leads to
\beq
    \frac{\d^2 \Phi}{\d z_{ip} \d z_{jq} } \, = \,
    \frac{ \al_i \al_j \tau_p \tau_q}{l_i (\tau) l_j (\tau) } \Phi
    \, = \, \frac{\d^2 \Phi}{\d z_{iq} \d z_{jp} }
    \label{1-32}
\eeq
which automatically derives the equation (\ref{1-2c}).

The integral $F(Z)$ in (\ref{1-4}) therefore indeed satisfies the defining
equations (\ref{1-2a})-(\ref{1-2c}) of the generalized hypergeometric functions on
$Gr ( k+1 , n+1 )$.
{\it The Grassmannian space $Gr ( k+1 , n+1 )$
is defined as a set of $(k+1)$-dimensional linear
subspaces in $(n+1)$-dimensional complex vector space ${\bf C}^{n+1}$.}
It is defined as
\beq
    Gr ( k+1 ,n+1 ) \, = \, \widetilde{Z} / GL (k+1 ,{\bf C} )
    \label{1-33}
\eeq
where $\widetilde{Z}$ is $(k+1)\times (n+1)$
complex matrices with $rank \widetilde{Z} = k+1$.
Consider some matrix $M$ and assume that there exists
a nonzero $r$-dimensional minor determinant of $M$. Then
the rank of $M$ is in general defined by the largest number of
such $r$'s. Thus $\widetilde{Z}$ is not exactly same as $Z$
defined in (\ref{1-20}). $\widetilde{Z}$ is more relaxed
since it allows some $(k+1)$-dimensional minor determinants vanish,
that is, $Z \subseteq \widetilde{Z}$.
In this sense $F(Z)$ is conventionally called
the generalized hypergeometric functions on $Gr (k+1,n+1)$
and we follow this convention in the present note.

\vskip 0.5cm \noindent
\underline{Non-projected formulation}

In terms of the homogeneous coordinate
$\tau = ( \tau_0 , \tau_1 , \cdots , \tau_k )$,
the homogeneous coordinates on $\cp^k$,
coordinates on ${\bf C}^{k}$ can be parametrized as
\beq
    t_1 = \frac{\tau_1}{\tau_0} ~, ~ t_1 = \frac{\tau_1}{\tau_0} ~,~
    \cdots ~, ~ t_k = \frac{\tau_k}{\tau_0}
    \label{1-34}
\eeq
For simplicity, we now fix $(z_{00} , z_{10} , \cdots , z_{n0} )^T$ at
$(1, 0, \cdots, 0)^T$, {\it i.e.},
\beq
    Z = \left(
          \begin{array}{ccccc}
            1 & z_{01} & z_{02} & \cdots & z_{0n} \\
            0 & z_{11} & z_{12} & \cdots & z_{1n} \\
            \vdots & \vdots & \vdots &  & \vdots \\
            0 & z_{k1} & z_{k2} & \cdots & z_{kn} \\
          \end{array}
        \right)
    \label{1-35}
\eeq
Then the integrand of $F(Z)$ can be expressed as
\beqar
    \Phi \om & = &
    \tau_{0}^{\al_0}  \prod_{j = 1}^{n}
    \left( \tau_0 z_{0j} + \tau_1 z_{1j} + \cdots + \tau_k z_{kj} \right)^{\al_j} \,
    \nonumber \\
    && ~~ \times
    \sum_{i = 0}^{k} ( -1 )^i \tau_i
    d \tau_0 \wedge d \tau_1 \wedge \cdots \wedge
    d \tau_{i-1} \wedge d \tau_{i + 1} \wedge \cdots \wedge d \tau_k
    \nonumber \\
    &=&
    \prod_{j = 1}^{n}
    \left(  z_{0j} + \frac{\tau_1}{\tau_0} z_{1j} + \cdots + \frac{\tau_k}{\tau_0} z_{kj} \right)^{\al_j}
    d \left( \frac{\tau_1}{\tau_0} \right) \wedge d \left( \frac{\tau_2}{\tau_0} \right)
    \wedge \cdots \wedge d \left( \frac{\tau_k}{\tau_0} \right)
    \nonumber \\
    & =& \widetilde{\Phi} \widetilde{\om}
    \label{1-36}
\eeqar
where we use (\ref{1-3b}) and define $\widetilde{\Phi}$, $\widetilde{\om}$ by
\beqar
    \widetilde{\Phi}  &=& \prod_{j=1}^{n} \widetilde{l}_j (t)^{\al_j}
    \label{1-37a} \\
    \widetilde{l}_j (t) &=& z_{0j} + t_1 z_{1j} + t_2 z_{2j} + \cdots + t_k z_{kj}
    ~~~~~ ( 1 \le j \le n )
    \label{1-37b} \\
    \widetilde{\om} &=& dt_1 \wedge dt_2 \wedge \cdots \wedge dt_k
    \label{1-37c}
\eeqar
The exponents $\al_j$ $(j= 1,2,\cdots, n)$ are also imposed to
the non-integer conditions $\al_j \not\in  {\bf Z} $
and $\al_1 + \al_2 + \cdots + \al_n \not\in  {\bf Z}$.
The multivalued function $\widetilde{\Phi}$ is now defined
in the following space
\beq
    \widetilde{X} \, = \, {\bf C}^k - \bigcup_{j= 1}^{n} \widetilde{{\cal H}}_j
    \label{1-38}
\eeq
where
\beq
    \widetilde{{\cal H}}_j \, = \,  \{ t \in {\bf C}^k \, ; ~ \widetilde{l}_j (t) = 0 \}
    \label{1-39}
\eeq
These are non-projected versions of (\ref{1-6}) and (\ref{1-7}).

As before, from $\widetilde{\Phi}$ we can define
rank-1 local systems $\widetilde{\cal L}$,
$\widetilde{\cal L}^{\vee}$ on $\widetilde{X}$, which lead
to the $k$-th homology and cohomology groups,
$H_k ( \widetilde{X}, \widetilde{\cal L}^{\vee})$
and $H^k ( \widetilde{X}, \widetilde{\cal L})$.
Then the integral over $\widetilde{\Phi} \widetilde{\om}$ is defined as
\beq
    F(Z) \, = \, \int_{\widetilde{\Del}}\widetilde{\Phi} \widetilde{\om}
    \label{1-40}
\eeq
where $[ \widetilde{\Del} ] = H_k ( \widetilde{X}, \widetilde{\cal L}^{\vee})$
and $[ \widetilde{\om} ] = H^k ( \widetilde{X}, \widetilde{\cal L})$.

In regard to the cohomology group $H^k ( \widetilde{X}, \widetilde{\cal L})$,
Aomoto shows the following theorem\footnote{Theorem 9.6.2 in \cite{Aomoto:1994bk}}:
\begin{enumerate}
  \item The dimension of $H^k ( \widetilde{X}, \widetilde{\cal L})$ is given
  by $\left(
       \begin{array}{c}
        \!\! n-1 \!\! \\
        \!\! k \!\! \\
       \end{array}
     \right)$.
  \item The basis of $H^k ( \widetilde{X}, \widetilde{\cal L})$ can be formed by
  $d \log \widetilde{l}_{j_1} \wedge d \log \widetilde{l}_{j_1} \wedge
  \cdots \wedge d \log \widetilde{l}_{j_k}$ where
  $1\le j_1 < j_2 < \cdots  < j_k \le n-1$.
\end{enumerate}
Correspondingly, the homology group
$H_k ( \widetilde{X}, \widetilde{\cal L}^{\vee})$ has dimension
$\left(
   \begin{array}{c}
    \!\! n-1 \!\! \\
    \!\! k \!\! \\
   \end{array}
\right)$ and its basis can be formed
finite regions bounded by $\widetilde{\cal H}_j$.
In terms of $\widetilde{l}_j$'s the basis
of $H^k ( \widetilde{X}, \widetilde{\cal L})$ can also be
chosen as \cite{Haraoka:2002bk}:
\beq
    \varphi_{j_1 j_2 \dots j_k} \, = \,
    d \log \frac{ \widetilde{l}_{j_1 + 1}}{ \widetilde{l}_{j_1}}
    \wedge
    d \log \frac{ \widetilde{l}_{j_2 + 1}}{ \widetilde{l}_{j_2}}
    \wedge
    \cdots
    \wedge
    d \log \frac{ \widetilde{l}_{j_k + 1}}{ \widetilde{l}_{j_k}}
    \label{1-41}
\eeq
where $1\le j_1 < j_2 < \cdots  < j_k \le n-1$.

\section{Generalized hypergeometric functions on $Gr ( 2, n + 1 )$}

In this section we consider a particular case of $Gr( 2, n+1)$.
The corresponding configuration space is simply given
by $n+1$ distinct points in $\cp^1$. This can be represented
by a $2 \times (n+1)$ matrix $Z$ any of whose 2-dimensional
minor determinants are nonzero.
Allowing the freedom of
coordinate transformations $GL(2, {\bf C})$ from the right
and scale transformations $H_2 = \diag(h_0, h_1)$, we can
uniquely parametrize $Z$ as
\beq
    Z =
    \left(
      \begin{array}{cccccc}
        1 & 0 & 1 & 1 & \cdots & 1 \\
        0 & 1 & -1 & -z_3 & \cdots & -z_{n} \\
      \end{array}
    \right)
    \label{2-1}
\eeq
where $z_i \ne 0 , 1, z_j$ ($i\ne j$, $3 \le i, j \le n$).
Thus we can regard $Z$ as
\beq
    Z \, \simeq \, \{ (z_3 , z_4 , \cdots , z_{n} ) \in {\bf C}^{n-2} \, | \,
    z_i \ne 0, 1, z_j ~ (i\ne j) \}
    \label{2-2}
\eeq
The three other points $(z_0 , z_1 , z_2 )$ can be fixed at $\{ 0, 1, \infty \}$
This agrees with the fact that the $GL(2, {\bf C})$
invariance fixes three points out of the $(n+1)$ distinct points in $\cp^1$.

In application of the previous section, we can carry out
a systematic formulation of the generalized hypergeometric functions
on $Gr(2, n+1)$ as follows.
We begin with a multivalued function of a form
\beq
    \Phi =   1^{\al_0} \cdot t^{\al_1} ( 1 -t)^{\al_2}
    (1 - z_3 t)^{\al_3} \cdots ( 1 - z_{n} t )^{\al_n} \, = \, \prod_{j=1}^{n} l_j (t)^{\al_j}
    \label{2-3}
\eeq
where
\beq
    l_0(t) = 1 \, , ~ l_1(t) = t \, , ~ l_2 (t) = 1 -t
    \, , ~ l_j (t) = 1 - z_{j} t ~~ (3 \le j \le n)
    \label{2-4}
\eeq
As in (\ref{1-5a}, \ref{1-5b}), the exponents obey the non-integer conditions
\beq
    \al_j \not\in {\bf Z} ~~ (0 \le j \le n)\, ,~~~  \sum_{j=0}^{n} \al_j = -2
    \label{2-5}
\eeq
As considered before, the latter condition applies to the expression (\ref{1-37c}),
that is, when $F(Z)$ is expressed as $F(Z) = \int_\Del \Phi dt$.
The defining space of $\Phi$ is given by
\beq
    X = \cp^1 - \{ 0 ,1, 1/ z_{3} , \cdots , 1/ z_{n} , \infty \}
    \label{2-6}
\eeq
From $\Phi$ we can determine a rank-1 local system ${\cal L}$ on $X$
and its dual local system ${\cal L}^{\vee}$.
Applying the result in (\ref{1-41}), the basis of the cohomology
group $H^1 ( X , {\cal L})$ is then given by
\beq
    d \log \frac{l_{j+1}}{l_j} ~~~~~~ (0 \le j \le n-1)
    \label{2-7}
\eeq
In the present case the basis of the homology group $H_1 ( X , {\cal L}^{\vee} )$
can be specified by a set of paths connecting the branch points.
For example, we can choose these by
\beq
    \Del_{\infty 0} \, , \,
    \Del_{01} \, , \, \Del_{1\frac{1}{z_3}} \, , \, \Del_{\frac{1}{z_3}\frac{1}{z_4} }
    \, , \, \cdots \, , \, \Del_{\frac{1}{z_{n-1}}\frac{1}{z_{n}} }
    \label{2-8}
\eeq
where $\Del_{pq}$ denotes a path on $\cp^1$ connecting branch points $p$ and $q$.
To summarize, for an element $\Del \in H_1 ( X , {\cal L}^{\vee} )$
associated with $\Phi$ of (\ref{2-3}),  we can define
a set of generalized hypergeometric functions on $Gr ( 2, n+1)$ as
\beq
    f_j ( Z ) \, = \, \int_\Del \Phi \, d \log \frac{l_{j+1}}{l_j}
    \label{2-9}
\eeq
where $0 \le j \le n-1$.
In the next section we consider the case of $n=3$, the simplest
case where only one variable exists, which corresponds to Gauss' hypergeometric function.

\section{Reduction to Gauss' hypergeometric function}
\noindent
\underline{Basics of Gauss' hypergeometric function}

We first review the basics of Gauss' hypergeometric function.
In power series, it is defined as
\beq
    F (a, b, c; z) \, = \, \sum_{n= 1}^{\infty}
    \frac{ (a)_n (b)_n }{ (c)_n \, n! } z^n
    \label{2-10}
\eeq
where $|z| < 1$, $c \not \in {\bf Z }_{\le 0}$ and
\beq
    (a)_{n} =
    \left\{
      \begin{array}{ll}
        1 & ~~ (n=1) \\
        a(a+1)(a+2) \cdots (a+n-1) & ~~(n \ge 1) \\
      \end{array}
    \right.
    \label{2-11}
\eeq
$F (a, b, c; z)$ satisfies the hypergeometric differential equation
\beq
    \left[
    \frac{d^2}{dz^2} + \left( \frac{c}{z} + \frac{a+b+1 - c}{z-1} \right) \frac{d}{d z}
    + \frac{ab}{z (z-1) }
    \right] F ( a, b,c; z)  \, = \, 0
    \label{2-12}
\eeq
Euler's integral formula for $F ( a, b,c; z)$ is written as
\beq
    F ( a, b,c; z) \, = \, \frac{\Ga ( c) }{\Ga( a ) \Ga (c - a)}
    \, \int_{0}^{1} t^{a-1} (1 -t )^{c-a-1} (1 - z t )^{-b} \, dt
    \label{2-13}
\eeq
where $| z | < 1 $ and $0 < \Re(a) < \Re(c ) $\footnote{
This condition can be relaxed to $a \not \in {\bf Z}$,
$c-a \not \in {\bf Z}$ by use of the well-known Pochhammer contour
in the integral (\ref{2-12}).
}.
$\Ga (a)$'s denote the Gamma functions
\beq
    \Gamma (a ) \, = \, \int_{0}^{\infty} e^{-t} t^{a -1} dt
    ~~~~~ ( \Re (a) > 0 )
    \label{2-13a}
\eeq
The second order differential equation (\ref{2-12}) has regular singularities
at $z = 0, 1, \infty$.
Two independent solutions around each singular point are expressed as
\beqar
    z = 0 ~: && \left\{
    \begin{array}{l}
      f_1 (z) \, = \, F(a,b,c; z) \\
      f_2 (z) \, = \, z^{1-c} F( a-c+1, b-c+1, 2-c;z)
    \end{array}
    \right.
    \label{2-14a} \\
    z = 1  ~: && \left\{
    \begin{array}{l}
      f_3 (z) \, = \, F(a,b,a+b-c+1; 1-z) \\
      f_4 (z) \, = \, (1-z)^{c-a-b} F( c-a, c-a, c-a-b+1;1-z)
    \end{array}
    \right.
    \label{2-14b} \\
    z = \infty ~: &&  \left\{
    \begin{array}{l}
      f_5 (z) \, = \, z^{-a} F(a,a-c+1,a-b+1; 1/z) \\
      f_6 (z) \, = \, z^{-b} F(b-c+1, b,b-a+1;1z)
    \end{array}
    \right.
    \label{2-14c}
\eeqar
where we assume $c \not \in {\bf Z}$,
$a+b-c \not \in {\bf Z}$
and $a-b \not \in {\bf Z}$
at $z= 0$, $z = 1$
and $z = \infty$, respectively.

\vskip 0.5cm \noindent
\underline{Reduction to Gauss' hypergeometric function 1: From defining equations}

From (\ref{2-13}) we find the relevant $2 \times 4$ matrix in a form of
\beq
    Z =
    \left(
      \begin{array}{cccc}
        1 & 0 & 1 & 1  \\
        0 & 1 & -1 & -z  \\
      \end{array}
    \right)
    \label{2-15}
\eeq
The set of equations (\ref{1-2a})-(\ref{1-2c})
then reduce to the followings:
\beqar
    ( \d_{00} + \d_{02} + \d_{03} ) F &=& -F
    \label{2-16a} \\
    ( \d_{11} - \d_{12} + z \d_z ) F &=& -F
    \label{2-16b} \\
    ( \d_{10} + \d_{12} - \d_{z} ) F &=& 0
    \label{2-16c} \\
    ( \d_{01} + \d_{02} - \d_{03} ) F &=& 0
    \label{2-16d} \\
    \d_{00} F &=& \al_0 F
    \label{2-17a} \\
    \d_{11} F &=& \al_1 F
    \label{2-17b} \\
    ( \d_{02} - \d_{12} ) F &=& \al_2 F
    \label{2-17c} \\
    ( \d_{03} + z \d_{z} ) F &=& \al_3 F
    \label{2-17d} \\
    -  \d_{z} \d_{02} F &= & \d_{12} \d_{03} F
    \label{2-18}
\eeqar
where $\d_{ij} = \frac{\d}{\d z_{ij}}$ and $\d_{13} = - \frac{\d}{\d z} = - \d_z$.
The last relation (\ref{2-18}) arises from (\ref{1-2c}); we here write
down the one that is nontrivial and involves $\d_z$.
Since the sum of (\ref{2-16a}) and (\ref{2-16b}) equals to
the sum of (\ref{2-17a})-(\ref{2-17d}), we can easily find
$\al_0 + \cdots + \al_3 = -2$ in accord with (\ref{2-5}).
The second order equation (\ref{2-18}) is then expressed as
\beq
    - \d_z ( \al_1 + \al_2 + 1 + z \d_z  ) F \, = \,
    ( \al_1 + 1 + z \d_z ) ( \al_3 - z \d_z ) F
    \label{2-19}
\eeq
This can also be written as
\beq
    \left[
    z (1-z) \d^{2}_{z} + \left( c - (a+ b+1) z \right) \d_z -ab
    \right] F \, = \, 0
    \label{2-20}
\eeq
where
\beqar
    a &=& \al_1 + 1
    \nonumber \\
    b &=& - \al_3
    \label{2-21} \\
    c &=& \al_1 + \al_2 + 2
    \nonumber
\eeqar
We can easily check that (\ref{2-20})
identifies with the hypergeometric differential equation (\ref{2-12}).

As seen in (\ref{2-1}), there exist multiple complex variables
for $n > 3$. In these cases reduction of the defining equations
(\ref{1-2a})-(\ref{1-2c}) can be carried out in principle but, unfortunately,
is not as straightforward as the case of $n=3$.

\vskip 0.5cm \noindent
\underline{Reduction to Gauss' hypergeometric function 2: Use of twisted cohomology}

The hypergeometric equation (\ref{2-12}) is a second order differential equation.
Setting $f_1 = F$, $f_2 =  \frac{z}{b} \frac{d}{dz} F$, we can express (\ref{2-12})
in a form of a first order Fuchsian differential equation \cite{Aomoto:1994bk}:
\beq
    \frac{d}{d z}
    \left(
      \begin{array}{c}
        f_1 \\
        f_2 \\
      \end{array}
    \right)
    =
    \left(
    \frac{A_0}{z} + \frac{A_1}{z-1}
    \right)
     \left(
      \begin{array}{c}
        f_1 \\
        f_2 \\
      \end{array}
    \right)
    \label{2-22}
\eeq
where
\beq
    A_0 =
    \left(
      \begin{array}{cc}
        0 & b \\
        0 & 1-c \\
      \end{array}
    \right),
    ~~
    A_1 =
    \left(
      \begin{array}{cc}
        0 & 0 \\
        -a & c-a-b-1 \\
      \end{array}
    \right)
    \label{2-23}
\eeq
Using the results (\ref{2-3})-(\ref{2-9}), we now
obtain other first order representations of
the hypergeometric function.

Let us start with a {\it non-projected} multivalued function
\beq
    \Phi \, = \, t^a (1-t)^{c-a} ( 1- zt )^{-b}
    \label{2-24}
\eeq
where
\beq
    a \, , \, c-a \, , \, -b \not \in {\bf Z}
    \label{2-25}
\eeq
$\Phi$ is defined on $X = \cp^1 - \{ 0,1,1/z , \infty \}$.
From these we can determine
a rank-1 local system ${\cal L}$ and its dual ${\cal L}^\vee$ on $X$.
Then, using (\ref{2-7}), we can obtain
a basis of the cohomology group $H^1 ( X , {\cal L})$ given by
the following set
\beqar
    \varphi_{\infty 0} &=& \frac{dt}{t}
    \label{2-26a} \\
    \varphi_{01} &=& \frac{dt}{t (1-t)}
    \label{2-26b} \\
    \varphi_{1 \frac{1}{z}} &=& \frac{(z-1) dt}{(1-t) ( 1 -zt)}
    \label{2-26c}
\eeqar
Similarly, from (\ref{2-8}) a basis of the homology group $H_1 ( X ,{\cal L}^\vee )$
is given by
\beq
    \{ \Del_{\infty 0} \, , \, \Del_{01} \, , \, \Del_{1 \frac{1}{z}}
    \}
    \label{2-27}
\eeq
In terms of these we can express Gauss' hypergeometric function as
\beq
    f_{01} (Z)
    \,  =  \,  \int_{\Del_{01}}  \Phi \varphi_{01}
    \, = \,
    \int_{0}^{1} t^{a-1} ( 1- t)^{c-a-1} ( 1 - z t)^{-b}  dt
    \label{2-28}
\eeq

The derivative of $f_{01} (Z) = f_{01}(z)$ with respect to $z$ is written as
\beq
    \frac{d}{d z} f_{01}(z) \, = \,
    \frac{d}{d z} \int_{\Del_{01}}  \Phi \varphi_{01}
    \, = \,
    \int_{\Del_{01}}  \Phi  \, \nabla_{\! z} \varphi_{01}
    \label{2-29}
\eeq
where
\beq
    \nabla_{\! z} \, = \, \d_z + \d_z \log \Phi
    \, = \, \d_z + \frac{bt}{1-zt}
    \label{2-30}
\eeq
Thus the derivative comes down to the computation of
$\nabla_{\! z} \varphi_{01}$; notice that the choice of a twisted cycle $\Del$
is irrelevant as far as the derivative itself is concerned.
In order to make sense of (\ref{2-29}) we should require
$\nabla_{\! z} \varphi_{01} \in H^1 ( X , {\cal L})$, that
is, it should be represented by a linear combinations of
(\ref{2-26a})-(\ref{2-26c}).
There is a caveat here, however. We know that
an element of $H^1 ( X , {\cal L})$ forms an equivalent class
as discussed earlier; see (\ref{1-8}) and (\ref{1-9}).
In the present case ($k=1$), $\al$ in (\ref{1-8}) is a 0-form
or a constant. So we can demand
\beq
    d \log \Phi
    \, = \,
    a \frac{dt}{t} - (c-a) \frac{dt}{1-t}
    + b \frac{z \, dt}{1-zt}
    \, \equiv \,
    0
    \label{2-31}
\eeq
in the computation of $\nabla_{\! z} \varphi_{01}$.
This means that the number of the base elements
can be reduced from 3 to 2. Namely, any elements of
$H^1 ( X , {\cal L})$ can be expressed by a combinations
of an arbitrary pair in (\ref{2-26a})-(\ref{2-26c})
under the condition (\ref{2-31}).
This explains the
numbering discrepancies between (\ref{1-41}) and
(\ref{2-7}) and agrees with the general result in the previous section that
the dimension of the cohomology group is given by
$\left(
   \begin{array}{c}
    \!\! n-1 \!\! \\
     \!\! k \!\! \\
   \end{array}
 \right) =
 \left(
   \begin{array}{c}
    \!\! 2 \!\! \\
    \!\! 1 \!\! \\
   \end{array}
 \right) = 2$.

Choosing the pair of $( \varphi_{01}, \varphi_{\infty 0} )$, we find
\beqar
    \nabla_{\! z} \varphi_{\infty 0} &=& \frac{b dt}{1-zt}
    \nonumber \\
    & \equiv &
    \frac{1}{z} \left( - a \frac{dt}{t} + (c-a) \frac{dt}{1-t} \right)
    \nonumber \\
    &=& \frac{c-a}{z} \varphi_{01} - \frac{c}{z} \varphi_{\infty 0}
    \label{2-32a} \\
    \nabla_{\! z} \varphi_{01} &=&
    \nabla_{\! z} \left( \varphi_{\infty 0} + \frac{dt}{1-t} \right)
    \nonumber \\
    &=&
    \nabla_{\! z} \varphi_{\infty 0} + \frac{b}{1-z} \left( \frac{dt}{1-t} - \frac{dt}{1-zt} \right)
    \nonumber \\
    &=&
    \frac{z}{z-1} \nabla_{\! z} \varphi_{\infty 0} - \frac{b}{z-1} ( \varphi_{01} - \varphi_{\infty 0} )
    \nonumber \\
    & \equiv &
    \frac{c-a-b}{z-1} \varphi_{01} + \frac{b-c}{z-1} \varphi_{\infty 0}
    \label{2-32b}
\eeqar
where notation $\equiv$ means the use of condition (\ref{2-31}).
Using (\ref{2-29}), we obtain a first order differential equation
\beq
    \frac{d}{dz}
    \left(
       \begin{array}{c}
         f_{01} \\
         f_{\infty 0} \\
       \end{array}
    \right)
    =
    \left(
        \frac{A^{(\infty 0)}_{0}}{z}   + \frac{A^{(\infty 0)}_{1}}{z -1 }
    \right)
    \left(
       \begin{array}{c}
         f_{01} \\
         f_{\infty 0} \\
       \end{array}
     \right)
    \label{2-33}
\eeq
where
\beq
    A^{(\infty 0)}_{0} =
    \left(
      \begin{array}{cc}
        0 & 0 \\
        c-a & -c \\
      \end{array}
    \right),
    ~~
    A^{(\infty 0)}_{1} =
    \left(
      \begin{array}{cc}
        c-a-b & b-c \\
        0 & 0 \\
      \end{array}
    \right)
    \label{2-34}
\eeq
Solving for $f_{01}$, we can easily confirm that (\ref{2-33}) leads
to Gauss' hypergeometric differential equation (\ref{2-12}).

Similarly, for the choice of
$( \varphi_{01}, \varphi_{1 \frac{1}{z} } )$ we find
\beqar
    \nabla_{\! z} \varphi_{01}
    &=&
    \frac{b}{z-1} \varphi_{1 \frac{1}{z} }
    \label{2-35a} \\
    \nabla_{\! z} \varphi_{1 \frac{1}{z} }
    & \equiv &
    \nabla_{\! z} \left( -\frac{a}{b} \frac{z-1}{z} \varphi_{01} +
    \frac{c-a}{b} \frac{z-1}{z} \frac{dt }{(1-t)^2} \right)
    \nonumber \\
    & \equiv &
    - \frac{a}{z} \varphi_{01} + \left( - \frac{c+1}{z} + \frac{c-a-b+1}{z-1} \right)
    \varphi_{1 \frac{1}{z} }
    \label{2-35b}
\eeqar
Notice that
$\varphi_{01}$ and $\varphi_{1 \frac{1}{z} }$ have the same factor
$(1-t)^{-1}$. This factor can be absorbed in the definition of $\Phi$ in (\ref{2-24}).
Thus, in applying the derivative formula (\ref{2-29}), we should replace $c$
by $c-1$. This leads to another first order differential equation
\beq
    \frac{d}{dz}
    \left(
       \begin{array}{c}
         f_{01} \\
         f_{1 \frac{1}{z}} \\
       \end{array}
    \right)
    =
    \left(
        \frac{A^{(1 \frac{1}{z})}_{0}}{z}   + \frac{A^{(1 \frac{1}{z})}_{1}}{z -1 }
    \right)
    \left(
       \begin{array}{c}
         f_{01} \\
         f_{1 \frac{1}{z}} \\
       \end{array}
     \right)
    \label{2-36}
\eeq
where
\beq
    A^{(1 \frac{1}{z})}_{0} =
    \left(
      \begin{array}{cc}
        0 & 0 \\
        -a & -c \\
      \end{array}
    \right),
    ~~
    A^{(1 \frac{1}{z})}_{1} =
    \left(
      \begin{array}{cc}
        0 & b \\
        0 & c-a-b \\
      \end{array}
    \right)
    \label{2-37}
\eeq
Solving for $f_{01}$, we can also check that
(\ref{2-36}) becomes Gauss' hypergeometric
differential equation (\ref{2-12}).

The representations (\ref{2-23}) and (\ref{2-34})
are obtained by Aomoto-Kita \cite{Aomoto:1994bk} and Haraoka
\cite{Haraoka:2002bk}, respectively.
The last one (\ref{2-37}) is not known in the literature
as far as the author notices.
Along the lines of the above derivation, we can also obtain
the Aomoto-Kita representation (\ref{2-23}) as follows.
We introduce a new one-form
\beq
    \widetilde{\varphi}_{1 \frac{1}{z}}
    \, =  \, \frac{z}{z-1} \varphi_{1 \frac{1}{z}}
    \, = \,  \frac{z \, dt}{ (1-t)(1-zt) }
    \label{2-38}
\eeq
The corresponding hypergeometric function is given by
$\widetilde{f}_{1 \frac{1}{z}} = \int_{\Del_{01}}
\Phi \widetilde{\varphi}_{1 \frac{1}{z}}$.
From (\ref{2-35a}) we can easily see $\nabla_{\! z} \varphi_{01} =
\frac{b}{z} \widetilde{\varphi}_{1 \frac{1}{z}}$. This is
consistent with the condition
$f_1 = F$, $f_2 =  \frac{z}{b} \frac{d}{dz} F$ in (\ref{2-22}).
Since $z$ is defined as $z \ne 0 ,1$,
$\frac{b}{z} \widetilde{\varphi}_{1 \frac{1}{z}}$ and
$\frac{b}{z-1} \varphi_{1 \frac{1}{z}}$ are equally well defined one-forms.
We can then choose
the pair $( \varphi_{01} , \widetilde{\varphi}_{1 \frac{1}{z}} )$
as a possible basis of the cohomology group.
The derivatives $\nabla_{\! z} \varphi_{01}$,
$\nabla_{\! z} \widetilde{\varphi}_{1 \frac{1}{z}} $ are calculated as
\beqar
    \nabla_{\! z} \varphi_{01}
    &=&
    \frac{b}{z} \widetilde{\varphi}_{1 \frac{1}{z}}
    \label{2-39a} \\
    \nabla_{\! z} \widetilde{\varphi}_{1 \frac{1}{z}}
    & \equiv &
    \nabla_{\! z} \left(
    - \frac{a}{b} \varphi_{01} + \frac{c-a}{b} \frac{dt}{ (1-t)^2 }
    \right)
    \nonumber \\
    & \equiv &
    - \frac{a}{z-1}  \varphi_{01} + \left( \frac{-c}{z} + \frac{c-a-b}{z-1}
    \right)\widetilde{\varphi}_{1 \frac{1}{z}}
    \label{2-39b}
\eeqar
where we use the relations
\beqar
    \frac{t \, dt }{(1-zt)( 1-t)} &=& \frac{1}{z-1} \left(
    \frac{dt}{1-zt} - \frac{dt}{1-t} \right)
    \label{2-40a} \\
    \frac{dt}{(1-t)^2} & \equiv & \frac{1}{c-a}
    \left( a \varphi_{01} + b \widetilde{\varphi}_{1 \frac{1}{z}} \right)
    \label{2-40b}
\eeqar
As before, $\varphi_{01}$ and $\widetilde{\varphi}_{1 \frac{1}{z}}$ have
the same factor $(1-t)^{-1}$. Thus, replacing $c$ by $c-1$, we obtain
a first order differential equation
\beq
    \frac{d}{dz}
    \left(
       \begin{array}{c}
         f_{01} \\
         \widetilde{f}_{1 \frac{1}{z}} \\
       \end{array}
    \right)
    =
    \left(
        \frac{\widetilde{A}^{(1 \frac{1}{z})}_{0}}{z}   + \frac{\widetilde{A}^{(1 \frac{1}{z})}_{1}}{z -1 }
    \right)
    \left(
       \begin{array}{c}
         f_{01} \\
         \widetilde{f}_{1 \frac{1}{z}} \\
       \end{array}
     \right)
    \label{2-41}
\eeq
where
$\widetilde{f}_{1 \frac{1}{z}} = \int_{\Del_{01}} \Phi \widetilde{\varphi}_{1 \frac{1}{z}}$
and
\beq
    \widetilde{A}^{(1 \frac{1}{z})}_{0} =
    \left(
      \begin{array}{cc}
        0 & b \\
        0 & 1-c \\
      \end{array}
    \right),
    ~~
    \widetilde{A}^{(1 \frac{1}{z})}_{1} =
    \left(
      \begin{array}{cc}
        0 & 0 \\
        -a & c-a-b-1 \\
      \end{array}
    \right)
    \label{2-42}
\eeq
We therefore reproduce the Aomoto-Kita representation (\ref{2-22}), (\ref{2-23})
by a systematic construction of first order representations of the hypergeometric
differential equation.

Lastly, we note that $\varphi_{\infty 0} = \frac{dt}{t}$ and $\varphi_{01} = \frac{dt}{t(1-t)}$
have the same factor $t^{-1}$ but we can not absorb this factor into $\Phi$.
This is because we can not obtain $dt$ as a base element of
$H^1 ( X , {\cal L})$ which is generically given in a form of
$d \log \frac{l_{j+1}}{l_j}$ as discussed in (\ref{2-7}).

\vskip 0.5cm \noindent
\underline{Reduction to Gauss' hypergeometric function 3: Permutation invariance}

The choice of twisted cycles or $\Del$'s is irrelevant in the above
derivations of the first order Fuchsian differential equations.
The hypergeometric differential equation is therefore
satisfied by a more general integral form, rather than (\ref{2-28}), {\it i.e.},
\beq
    f_{01}^{(\Del_{pq})} (z) \, = \,
    \int_{\Del_{pq}} \Phi \varphi_{01}
    \, = \, \int_{p}^{q} t^{a-1} ( 1- t)^{c-a-1} ( 1 - z t)^{-b}
    dt
    \label{2-43}
\eeq
where $(p, q)$ represents an arbitrary pair among the four branch points
$p, q \in \{ 0, 1, 1/z , \infty \}$.
This means that we can impose {\it permutation invariance} on
the branch points. $\Del_{pq}$ is then given by the following set of
twisted cycles:
\beq
    \Del_{pq} \, = \,
    \{  \Del_{\infty 0} \, , \, \Del_{01} \, , \, \Del_{1 \frac{1}{z}}
    \, , \,  \Del_{1 \infty} \, , \, \Del_{\frac{1}{z} \infty} \, , \, \Del_{0 \frac{1}{z}}
    \}
    \label{2-44}
\eeq
so that the number of elements becomes
$\left(
   \begin{array}{c}
    \!\! 4 \!\! \\
    \!\! 2 \!\! \\
   \end{array}
\right) = 6$. Correspondingly, the base elements of the cohomology group also include
\beqar
    \varphi_{1 \infty } &=& \frac{dt}{1-t}
    \label{2-45a} \\
    \varphi_{\frac{1}{z} \infty} &=& \frac{z \, dt}{1-zt}
    \label{2-45b} \\
    \varphi_{0 \frac{1}{z}} &=& \frac{ dt}{t ( 1 -zt)}
    \label{2-45c}
\eeqar
besides (\ref{2-26a})-(\ref{2-26c}).
It is known that $f_{01}^{(\Del_{pq})} (z)$ are related to
the local solutions $f_i (z)$ $(i= 1,2, \cdots , 6)$ in (\ref{2-14a})-(\ref{2-14c}) by
\beqar
      f_{01}^{(\Del_{01})} (z) & = & B( a, c-a) f_1 (z)
      \label{2-46a} \\
      f_{01}^{ ( \Del_{\frac{1}{z} \infty} )} (z) & = & e^{i \pi (a+b-c+1)} B( b-c+1 , 1-b) f_2 (z)
      \label{2-46b} \\
      f_{01}^{(\Del_{\infty 0})} (z) & = & e^{i \pi (1-a)} B( a , b-c+1) f_3 (z)
      \label{2-46c} \\
      f_{01}^{(\Del_{1 \frac{1}{z}})} (z) & = & e^{i \pi (a-c+1)} B( c-a , 1-b) f_4 (z)
      \label{2-46d} \\
      f_{01}^{(\Del_{1 \frac{1}{z}})} (z) & = & B( a , 1-b) f_5 (z)
      \label{2-46e} \\
      f_{01}^{(\Del_{1 \infty})} (z) & = & e^{- i \pi (a+b-c+1)} B( b-c+1 , c-a) f_6 (z)
      \label{2-46f}
\eeqar
where $B(a, b)$ is the beta function
\beq
    B (a, b) \, = \, \frac{\Ga (a) \Ga (b)}{\Ga(a+b)}
    \, = \, \int_{0}^{1} t^{a-1} (1-t)^{b-1} dt
    ~~~~~ ( \Re (a ) > 0 \, , \, \Re(b) > 0 )
    \label{2-47}
\eeq
(For derivations and details of these relations, see \cite{Haraoka:2002bk}.)

The relevant configuration space represented by $Z$ is given by
$Gr (2, 4) / \S_4$ where $\S_4$ denotes the rank-4 symmetry group.
The permutation invariance can also be confirmed by
deriving another set of the first order differential equations
with the choice of $\varphi_{01}$ and one of (\ref{2-45a})-(\ref{2-45c}).
This is what we will present in the following.

For the choice of
$( \varphi_{01}, \varphi_{1 \infty} )$ we find
\beqar
    \nabla_{\! z} \varphi_{1 \infty}
    &=&
    \frac{bt}{1-zt} \frac{dt}{(1-t)}
    \nonumber \\
    & \equiv &
    - \frac{a}{z(z-1)} \varphi_{01} + \left( \frac{c}{z(z-1)} - \frac{b}{z-1} \right)
    \varphi_{1 \infty}
    \label{2-48a} \\
    \nabla_{\! z} \varphi_{01 }
    &=& \nabla_{\! z} \frac{dt}{t} + \nabla_{\! z} \varphi_{1 \infty}
    \nonumber \\
    & \equiv &
    - \frac{a}{z-1} \varphi_{01} +  \frac{c-b}{z-1}  \varphi_{1 \infty }
    \label{2-48b}
\eeqar
The corresponding differential equation is then expressed as
\beq
    \frac{d}{dz}
    \left(
       \begin{array}{c}
         f_{01} \\
         f_{1 \infty} \\
       \end{array}
    \right)
    =
    \left(
        \frac{A^{(1 \infty)}_{0}}{z}   + \frac{A^{(1 \infty)}_{1}}{z -1 }
    \right)
    \left(
       \begin{array}{c}
         f_{01} \\
         f_{1 \infty} \\
       \end{array}
     \right)
    \label{2-49}
\eeq
where
\beq
    A^{(1 \infty)}_{0} =
    \left(
      \begin{array}{cc}
        0 & 0 \\
        a & -c \\
      \end{array}
    \right),
    ~~
    A^{(1 \infty)}_{1} =
    \left(
      \begin{array}{cc}
        -a & c-b \\
        -a & c-b \\
      \end{array}
    \right)
    \label{2-50}
\eeq
Solving for $f_{01}$, we can check that
(\ref{2-49}) indeed becomes Gauss' hypergeometric differential equation (\ref{2-12}).

Similarly, for $( \varphi_{01}, \varphi_{\frac{1}{z} \infty} )$ we find
\beqar
    \nabla_{\! z} \varphi_{01 }
    &=& \frac{b \, dt}{(1-zt)(1-t)}
    \nonumber \\
    & \equiv &
    - \frac{1}{z-1} \frac{ab}{c} \varphi_{01} -  \frac{1}{z-1} \frac{b}{c} (b-c)
    \varphi_{\frac{1}{z} \infty }
    \label{2-51a} \\
    \nabla_{\! z} \varphi_{\frac{1}{z} \infty}
    & \equiv &
    \nabla_{\! z} \left( -\frac{a}{b} \varphi_{01} + \frac{c}{b} \frac{dt}{1-t} \right)
    \nonumber \\
    & \equiv &
    \frac{1}{z-1} \frac{a}{c} (a -c) \varphi_{01} + \left( \frac{1}{z-1} \frac{1}{c} (b-c)(a-c)
    - \frac{c}{z} \right)
    \varphi_{\frac{1}{z} \infty}
    \label{2-51b}
\eeqar
The first order differential equation is then expressed as
\beq
    \frac{d}{dz}
    \left(
       \begin{array}{c}
         f_{01} \\
         f_{\frac{1}{z} \infty} \\
       \end{array}
    \right)
    =
    \left(
        \frac{A^{(\frac{1}{z} \infty)}_{0}}{z}   + \frac{A^{(\frac{1}{z} \infty)}_{1}}{z -1 }
    \right)
    \left(
       \begin{array}{c}
         f_{01} \\
         f_{\frac{1}{z} \infty} \\
       \end{array}
     \right)
    \label{2-52}
\eeq
where
\beq
    A^{(\frac{1}{z} \infty)}_{0} =
    \left(
      \begin{array}{cc}
        0 & 0 \\
        0 & -c \\
      \end{array}
    \right),
    ~~
    A^{(\frac{1}{z} \infty)}_{1} =
    \left(
      \begin{array}{cc}
        -\frac{ab}{c} & - \frac{b}{c}(b-c) \\
        \frac{a}{c}(a-c) & \frac{1}{c} (b-c)(a-c) \\
      \end{array}
    \right)
    \label{2-53}
\eeq
We can check that (\ref{2-52}) reduces to the hypergeometric
differential equation for $f_{01}$.

Lastly, for $(\varphi_{01} , \varphi_{0 \frac{1}{z}} )$ we find
\beqar
    \nabla_{\! z} \varphi_{01 }
    &=& \frac{b \, dt}{(1-zt)(1-t)}
    \nonumber \\
    & = &
    - \frac{b}{z-1} \left( \varphi_{01} -   \varphi_{0 \frac{1}{z} } \right)
    \label{2-54a} \\
    \nabla_{\! z} \varphi_{0 \frac{1}{z}}
    & \equiv &
    \frac{b \, dt}{ ( 1-zt)^2 }
    \nonumber \\
    & \equiv &
    - \frac{c-a}{z(z-1)}  \varphi_{01} + \frac{c - az}{z(z-1)} \varphi_{0 \frac{1}{z}}
    \label{2-54b}
\eeqar
The corresponding differential equation becomes
\beq
    \frac{d}{dz}
    \left(
       \begin{array}{c}
         f_{01} \\
         f_{0 \frac{1}{z}} \\
       \end{array}
    \right)
    =
    \left(
        \frac{A^{(0 \frac{1}{z})}_{0}}{z}   + \frac{A^{(0 \frac{1}{z})}_{1}}{z -1 }
    \right)
    \left(
       \begin{array}{c}
         f_{01} \\
         f_{0 \frac{1}{z}} \\
       \end{array}
     \right)
    \label{2-55}
\eeq
where
\beq
    A^{(0 \frac{1}{z})}_{0} =
    \left(
      \begin{array}{cc}
        0 & 0 \\
        c-a & -c \\
      \end{array}
    \right),
    ~~
    A^{(0 \frac{1}{z})}_{1} =
    \left(
      \begin{array}{cc}
        -b & b \\
        -c+a & c-a \\
      \end{array}
    \right)
    \label{2-56}
\eeq
We can check that (\ref{2-55}) reduces to the hypergeometric
differential equation for $f_{01}$ as well.

As in the case of (\ref{2-38}), it is tempting to
think of $\widetilde{\varphi}_{ \frac{1}{z} \infty}
=  \frac{z-1}{z} \varphi_{\frac{1}{z} \infty}
=  \frac{(z-1) \, dt}{ 1-zt }$.
But, with $\varphi_{01}$ and $\widetilde{\varphi}_{ \frac{1}{z} \infty}$,
it is not feasible to obtain a first order
differential equation in the form of (\ref{2-52})
which leads to the hypergeometric differential equation.
This is because, if expanded in $\varphi_{01}$ and $\widetilde{\varphi}_{ \frac{1}{z} \infty}$,
the $z$-dependence of the derivatives
$\nabla_{\! z}  \varphi_{01}$ and $\nabla_{\! z}  \widetilde{\varphi}_{ \frac{1}{z} \infty}$,
can not be written in terms of $\frac{1}{z}$ or $\frac{1}{z-1}$.

\vskip 0.5cm \noindent
\underline{Summary}

In this section we carry out a systematic derivation
of first order representations of the hypergeometric differential
equation by use of twisted cohomology as the simplest
reduction of Aomoto's generalized hypergeometric function.
The first order equations are generically expressed as
\beq
    \frac{d}{d z}
    \left(
      \begin{array}{c}
        f_{01} \\
        f_{pq} \\
      \end{array}
    \right)
    =
    \left(
    \frac{A^{(pq)}_{0}}{z} + \frac{A^{(pq)}_{1}}{z-1}
    \right)
     \left(
      \begin{array}{c}
        f_{01} \\
        f_{pq} \\
      \end{array}
    \right)
    = \, A^{(pq) }_{01}
     \left(
      \begin{array}{c}
        f_{01} \\
        f_{pq} \\
      \end{array}
    \right)
    \label{2-57}
\eeq
where $(pq)$ denotes a pair of four branch points
$\{ 0, 1 , 1/z , \infty \}$ in $\Phi = t^a (1-t)^{c-a} (1-zt)^{-b}$.
A list of the $(2 \times 2)$ matrices $A^{(pq)}_{01}$
obtained in this section is given by the following:
\beqar
    A_{01}^{( \infty 0)} &=&
    \left(
      \begin{array}{cc}
        \frac{c-a-b}{z-1} & \frac{b-c}{z-1} \\
        \frac{c-a}{z} & - \frac{c}{z} \\
      \end{array}
    \right)
    \label{2-58a} \\
    A_{01}^{(1 \frac{1}{z})} &=&
    \left(
      \begin{array}{cc}
        0 & \frac{b}{z-1} \\
        - \frac{a}{z} & - \frac{c}{z} + \frac{c-a-b}{z-1} \\
      \end{array}
    \right)
    \label{2-58b} \\
    \widetilde{A}_{01}^{(1 \frac{1}{z})} &=&
    \left(
      \begin{array}{cc}
        0 & \frac{b}{z} \\
        - \frac{a}{z-1} & - \frac{c-1}{z} + \frac{c-a-b-1}{z-1} \\
      \end{array}
    \right)
    \label{2-58c} \\
    A_{01}^{(1 \infty )} &=&
    \left(
      \begin{array}{cc}
        -\frac{a}{z-1} & \frac{c-b}{z-1} \\
        \frac{a}{z} - \frac{a}{z-1} & - \frac{c}{z} + \frac{c-b}{z-1} \\
      \end{array}
    \right)
    \label{2-58d} \\
    A_{01}^{( \frac{1}{z} \infty )} &=&
    \left(
      \begin{array}{cc}
        - \frac{1}{z-1} \frac{ab}{c} & - \frac{1}{z-1} \frac{b}{c} (b-c)  \\
        \frac{1}{z-1} \frac{a}{c} (a-c) & - \frac{c}{z} + \frac{1}{z-1} \frac{1}{c}(b-c) (a-c) \\
      \end{array}
    \right)
    \label{2-58e} \\
    A_{01}^{( 0 \frac{1}{z})} & =&
    \left(
      \begin{array}{cc}
        - \frac{b}{z-1} & \frac{b}{z-1} \\
        \frac{c-a}{z} - \frac{c-a}{z-1} & - \frac{c-1}{z} + \frac{c-a}{z-1} \\
      \end{array}
    \right)
    \label{2-58f}
\eeqar
where we include the Aomoto-Kita representation
$\widetilde{A}_{01}^{(1 \frac{1}{z})}$. As far as the author notices,
these expressions except (\ref{2-58a}, \ref{2-58c}) are new
for the description of the hypergeometric differential equation.
A common feature among these matrices is that the determinant is
identical:
\beq
    \det A^{(pq)}_{01} \, = \, \frac{ab}{z(z-1)}
    \label{2-59}
\eeq
In terms of the first order differential equation (\ref{2-57}),
this means that the action of the derivative on the basis
$
     \left(
      \begin{array}{c}
        f_{01} \\
        f_{pq} \\
      \end{array}
    \right)
$
of the cohomology group $H^1 ( X , {\cal L} )$
can be represented by a generator of the $SL(2, {\bf C})$ algebra.
In other words, a change of the bases
is governed by the $SL(2, {\bf C})$ symmetry.
The $SL(2, {\bf C})$ invariance corresponds to
the global conformal symmetry for holomorphic functions
on $\cp^1$. In the present case we start from
the holomorphic multivalued function $\Phi$ in (\ref{2-24}) which
is defined on $X = \cp^1 - \{ 0 , 1, 1/z , \infty \}$.
The result (\ref{2-59}) is thus natural in concept but nontrivial in practice because
the equivalence condition $d \log \Phi \equiv 0$ in (\ref{2-31})
is implicitly embedded into the expressions (\ref{2-58a})-(\ref{2-58f}).

\section{Integral representations of the KZ solutions}

Having investigated thoroughly reduction of
Aomoto's generalized hypergeometric function
to Gauss' original hypergeometric function,
we now consider a physical problem in relation to the above results.
As mentioned in the introduction, solutions of the Knizhnik-Zamolodchikov (KZ) equation
are known to be related to the generalized hypergeometric
functions. In this section we shed light on this relation along
the lines of arguments on cohomology and homology.

\vskip 0.5cm \noindent
\underline{The KZ equation}

We first review basic properties of the Knizhnik-Zamolodchikov (KZ) equation, following
the description in \cite{Kohno:2002bk}.
The KZ equation is defined by
\beq
    \frac{\d \Psi }{ \d z_i} =  \frac{1}{\kappa}
    \sum_{j \ne i}^{n} \frac{\Om_{ij} \Psi}{z_i - z_j}  ~~~~~ ( 1 \le i, j \le n)
    \label{3-1}
\eeq
where $\Psi = \Psi (z_1 , z_2 , \cdots , z_n)$ is a function of
$n$ complex variables, $z_i$ ($i = 1, 2, \cdots, n$),
corresponding to a correlation function in the Wess-Zumino-Witten (WZW) model.
$\kappa$ is parametrized as
\beq
    \kappa \, = \, l + h^{\vee}
    \label{3-2}
\eeq
where $l$ is the level number and
$h^{\vee}$ is the dual Coexter number of the Lie algebra $\mathfrak{g}$ associated to the WZW model.
The function $\Psi$ is defined on a space
\beq
    X_n \, = \, {\bf C}^n  - \bigcup_{ i < j} {\cal H}_{ij}
    \label{3-3}
\eeq
where ${\cal H}_{i j}$ is denotes a hyperplane in ${\bf C}^n$ defined
by $z_i - z_j = 0$:
\beq
    {\cal H}_{i j} \, = \, \{ (z_1 , \cdots , z_n ) \in {\bf C}^n
     \, ; ~  z_i - z_j = 0 ~~(i \ne j)\}
    \label{3-4}
\eeq
Quantum theoretically, the function $\Psi$ should be
evaluated as a vacuum expectation value
of operators acting on the Hilbert space
\beq
    V^{\otimes n} = V_1 \otimes V_2 \otimes \cdots \otimes V_n
    \label{3-5}
\eeq
where $V_i$ ($i = 1,2, \cdots, n$) denotes a Fock space for a particle
labeled by $i$.
$\Om_{ij}$'s in the KZ equation are bialgebraic operators acting on
the $(i, j)$ entries of the Hilbert space $V^{\otimes n}$.
These operators satisfy {\it the infinitesimal braid relations}:
\beqar
    \left[ \Om_{ij} , \Om_{kl} \right] &=& 0  ~~~~~ \mbox{($i,j,k,l$ are distinct)}
    \label{3-6a} \\
    \left[ \Om_{ij} + \Om_{jk} , \Om_{ik} \right] &=& 0  ~~~~~ \mbox{($i,j,k$ are distinct)}
    \label{3-6b}
\eeqar

It is well-known that the KZ equation is invariant under
the global $SL(2 ,{\bf C})$ symmetry or the conformal transformations.
It is then natural to consider each variable $z_i$ on $\cp^1$ rather than
on ${\bf C}$. In practice, this means that we can add
an extra variable $z_0 = \infty$ in the definitions of
$\Psi$, that is, $\Psi (z_1 , z_2 , \cdots , z_n)
\longrightarrow \Psi ( z_0 , z_1 , z_2 , \cdots , z_n)$.
In general, the solutions of the KZ equation (\ref{3-1})
give $(n+1)$-point correlation functions on $\cp^1$
in the WZW model, which we call $(n+1)$-point KZ solutions.
Fixing $(z_1 , z_2 ) =( 0 , 1)$, we can then identify
the configuration space (\ref{3-3}) as
the $2 \times (n+1)$ matrix $Z$ in (\ref{2-1}) defined
for the generalized hypergeometric function on $Gr(2, n+1)$.

In order to relate $\Psi$ to the generalized hypergeometric functions,
we need to determine a multivalued function analogous to
$\Phi$ in (\ref{1-5a}).
From the defining space (\ref{3-3}) we find that the relevant
multivalued function is given by
\beq
    \Phi_0 \,  = \, \prod_{1 \le i < j \le n} ( z_i  - z_j )^{ \frac{1}{\ka} \Om_{ij}}
    \label{3-9}
\eeq
In order to interpret $\Phi_0$ as a function, we
need to specify the Lie algebra $\mathfrak{g}$ for
$\Om_{ij}$ and relevant actions to the vacuum state.
{\it We shall not specify these algebraic properties here
and interpret $\Om_{ij}$ as $\bra \Om_{ij} \ket$,
the vacuum expectation value of $\Om_{ij}$, for the moment.}
We will specify the algebraic structure to $SL(2, {\bf C})$ in the next section.
In analogy to (\ref{1-10}) or (\ref{1-13}), we now
consider the following covariant derivatives
\beq
    D \, = \,  d  - d \log \Phi_0  \, = \,
    d - \frac{1}{\ka} \sum_{1 \le i < j \le n} \Om_{ij} \frac{ d (z_i  - z_j )}{z_i  - z_j}
    \label{3-10}
\eeq
General solutions of $D f  = 0$ are given by
\beq
    f \,  = \,  \la \Phi_0 \, = \, \la \prod_{1 \le i < j \le n}
    ( z_i  - z_j )^{ \frac{1}{\ka} \Om_{ij}} ~~~~~~~ (\la \in {\bf C}^{\times} )
    \label{3-11}
\eeq
As before, analytic continuation of these solutions forms
a fundamental homotopy group of a closed path in $X_n$.
The representation of $\Pi_1 (X_n )$ gives the monodromy representation
of the differential equation $D f = 0$.

We now notice that in a differential form
the KZ equation (\ref{3-1}) can be expressed as
\beq
    D \Psi \, = \, (d - \Om) \Psi \, = \, 0
    \label{3-12}
\eeq
where
\beqar
    \Om &=&  \frac{1}{\kappa} \sum_{1 \le i < j \le n} \Om_{ij} \, \om_{ij}
    \label{3-13} \\
    \om_{ij} &=& d \log (z_i - z_j) = \frac{ d z_i - d z_j}{z_i - z_j}
    \label{3-14}
\eeqar
Notice that from the result in section 2 we find that $\om_{ij}$'s form
a basis of the cohomology group of $X_n$. These satisfy the identity
\beq
    \om_{ij} \wedge \om_{jk} + \om_{jk} \wedge \om_{ik} + \om_{ik} \wedge \om_{ij} = 0
    ~~~~~~~ (i < j < k)
    \label{3-15}
\eeq
From (\ref{3-6a}), (\ref{3-6b}) and (\ref{3-15}) we can show the flatness of $\Om$, {\it i.e.},
\beq
    d \Om - \Om \wedge \Om \, =  \, 0
    \label{3-16}
\eeq
$\Om$ is called the KZ connection one-form.

Imposing permutation invariance on $X_n$, we can in fact express the defining
space as
\beq
    \C = \frac{X_n }{ \S_n }
    \label{3-17}
\eeq
where $\S_n$ is the rank-$n$ symmetric group.
The fundamental homotopy group of $\C$ is given by the the braid group
\beq
    \Pi_1 ( \C ) \, = \, \B_n
    \label{3-18}
\eeq
The braid group $\B_n$ has generators, $b_1 , b_2 , \cdots ,
b_{n-1}$, and they satisfy the following relations
\beqar
    b_i b_{i+1} b_i &=& b_{i+1} b_i b_{i+1} ~~~~ (|i - j| = 1 )
    \label{3-19a} \\
    b_i b_j &=& b_j b_i  ~~~~~~~~~~~~ (|i - j| > 1 )
    \label{3-19b}
\eeqar
where we identify $b_n$ with $b_1$.
One of the main results in the KZ equation is that the monodromy
representation of the KZ equation can be given by the braid group.

\vskip 0.5cm \noindent
\underline{Relation to 4-point KZ solutions}

For $n=3$ we can check that the KZ equation (\ref{3-1})
has a solution of the form \cite{Kohno:2002bk}:
\beq
    \Psi ( z_1 , z_2 , z_3 ) \, = \,
    (z_3 - z_1 )^{\frac{1}{\ka} ( \Om_{12} + \Om_{13} + \Om_{23} ) }
    \, G \left(
    \frac{z_2 - z_1}{z_3 - z_1}
    \right)
    \label{3-20}
\eeq
where $G(z)$ satisfies the differential equation
\beq
    \frac{d}{d z} G (z) \, = \,
    \frac{1}{\ka} \left(
    \frac{ \Om_{12}}{z} + \frac{ \Om_{23} }{ z - 1}
    \right) G(z)
    \label{3-21}
\eeq
This equation is equivalent in structure to the first order differential equation (\ref{2-57}).
The solution $G(z)$ thus corresponds to Gauss' hypergeometric
function if $\Om_{12}$ and $\Om_{23}$ are represented by $2 \times 2$ matrices.
In what follows we shall clarify this statement
by constructing hypergeometric-type integral representations of the
4-point KZ solutions.

As before, $G(z)$ has singular points at $0, 1, \infty$.
Thus, evaluated on the Riemann sphere $\cp^1$, the solution (\ref{3-20})
can be interpreted as a 4-point solution.
Thanks to the $SL(2, {\bf C})$ invariance, without losing
generality, we can fix the three points $(0, 1, \infty )$.
For example, choosing $(z_0 ,  z_1 , z_2 , z_3 )  \rightarrow ( \infty , 0 , z ,1 )$,
we find
\beq
    \Psi (z_0 ,  z_1 , z_2 , z_3 ) \,  \rightarrow \, \Psi ( \infty , 0 , z ,1 )
    \, = \, G (z)
    \label{3-22}
\eeq
where we omit the $z$-independent factor
$(z_3 - z_1)^{\frac{1}{\ka} ( \Om_{12} + \Om_{13} + \Om_{23} )} \rightarrow
e^{i 2 \pi m \frac{1}{\ka} ( \Om_{12} + \Om_{13} + \Om_{23} ) }$ with
$m \in {\bf Z}$. When evaluated as a vacuum expectation value (vev),
this factor enters nontrivially into the solution $\Psi$.
If the exponent is evaluated as an integer we can not
properly define $\Phi$ or $\Phi_0$ since these function can
be multiplied by $1 = e^{i 2 \pi m}$ for any times.
Thus we can naturally demand non-integer conditions for
the vev of these exponents:
\beq
    \left\langle \frac{1}{\ka}\Om_{12} \right\rangle ,
    \left\langle \frac{1}{\ka}\Om_{13} \right\rangle ,
    \left\langle \frac{1}{\ka}\Om_{23} \right\rangle ,
    \left\langle \frac{1}{\ka} ( \Om_{12} + \Om_{13} + \Om_{23} ) \right\rangle
    \, \not \in {\bf Z}
    \label{3-23}
\eeq
These conditions are analogous to the non-integer conditions we have
considered for the exponents $\al_j$, say, in (\ref{1-37a}).
As mentioned earlier, we will omit the brackets in the following expressions.
In terms of the covariant derivative (\ref{3-10}), we find
\beq
    D_{z_2} = \d_{z_2} - \d_{z_2 } \log \Phi_0
    \, \rightarrow \, D_z =
    \d_z - \frac{1}{\ka} \left(
    \frac{ \Om_{12}  }{ z} + \frac{ \Om_{23} }{z -1}
    \right)
    \label{3-24}
\eeq
In this case the KZ equation $D_{z} \Psi = 0$ directly
reduces to the differential equation (\ref{3-21}).
In general, the reduction is not so simple but, as expected,
we may relate the KZ equation $D_z \Phi  = 0$ to
the equation (\ref{3-21}).
For example, take the previous choice
$(z_0 ,  z_1 , z_2 , z_3 )  \rightarrow ( \infty , 0 , 1 , 1/z )$, which leads to
the following parametrization:
\beqar
    \Psi (z_0 ,  z_1 , z_2 , z_3 ) & \rightarrow & \Psi ( \infty , 0 ,1, 1/ z )
    \, = \, z^{-\frac{1}{\ka} ( \Om_{12} + \Om_{13} + \Om_{23} )} G (z)
    \label{3-25a} \\
    \Phi_0 &=& ( -1 )^{\frac{1}{\ka} \Om_{12}}( 1-1/z )^{\frac{1}{\ka} \Om_{23}}
    ( -1/z )^{\frac{1}{\ka} \Om_{13}}
    \label{3-25b} \\
    D_{1/ z_3} = \d_{1/ z_3}  - \d_{1/ z_3} \log \Phi_0
    & \rightarrow & D_z  =
    \d_z - \frac{1}{\ka} \left(
    - \frac{  \Om_{23} + \Om_{13} }{ z } + \frac{ \Om_{23} }{ z-1}
    \right)
    \label{3-25c} \\
    D_{z} \Psi  &=& \!
    z^{-\frac{1}{\ka} ( \Om_{12} + \Om_{13} + \Om_{23} )}
    \left[ D_z - \frac{1}{\ka} \left(
    \frac{ \Om_{12}  }{ z} + \frac{ \Om_{23} }{z -1}
    \right) \right] G(z) ~
    \label{3-25d} \\
    d \log \Phi_0 &=& \left(
    - \frac{  \Om_{23} + \Om_{13} }{ z } + \frac{ \Om_{23} }{ z-1}
    \right) dz \, \equiv \, 0
    \label{3-25e}
\eeqar
The last relation arises from the equivalent relations that are associated
to the cohomology group $H^1 (X_3 , {\cal L}_0 )$ where ${\cal L}_0$
is a rank-1 local system determined by $\Phi_0$.
Considering in one-form, we can express $D_z \Psi dz$ as
\beq
    D_z \Psi  dz \equiv z^{-\frac{1}{\ka} ( \Om_{12} + \Om_{13} + \Om_{23} )}
    \left[ \d_z - \frac{1}{\ka} \left(
    \frac{ \Om_{12}  }{ z} + \frac{ \Om_{23} }{z -1}
    \right) \right] G(z) dz
    \label{3-26}
\eeq
The KZ equation $D_z \Psi dz = 0$ then reduces to the differential equation
for $G(z)$ in (\ref{3-21}).
In the above analysis the concept of cohomology plays
an essential role to replace the covariant derivative $D_z$
by the ordinary derivative $\d_z$ in (\ref{3-25d}) and (\ref{3-26}).
In the language of gauge theory the replacement can be implemented
by taking a pure gauge of the KZ connection one-form.

Essence of the integral representation lies in the choice of
multivalued function. The choice of $\Phi_0$ in (\ref{3-25b}) is, however, not
appropriate to derive Gauss' hypergeometric function
as $\Phi_0$ does not contain the integral parameter $t$.
To obtain Gauss' hypergeometric function, we need to incorporate
the following multivalued function
\beq
    \Phi_t \, = \, t^{\al_1} ( 1- t )^{\al_2} ( 1-zt )^{\al_3}
    \label{3-27}
\eeq
which is defined on $X = \cp^1 - \{ \infty , 0 , 1 , 1/z \}$.
Note that the choice of $X_3$ automatically determines $X$.
As discussed earlier, $X_n$ is represented by distinct  $(n+1)$ points in $\cp^1$.
The physical configuration
space $\C = X_n / \S_n$ imposes permutation (or bosonic) invariance
on these so that it gives rise to distinct {\it ordered}
$(n+1)$ points in $\cp^1$.

Another issue on $\Phi_0$ is that it
is an algebraic function because of the exponent $\frac{1}{\ka}\Om_{ij}$.
A rank-1 local system determined by $\Phi_0$ is therefore an algebraic one.
To incorporate $\Phi_t$ into $\Phi_0$ we then need to demand
that the exponents $\al_i$ of $\Phi_t$ be algebraic.
To be concrete, we set
\beq
    \al_i = - \frac{1}{\ka} \Om_i ~~~~~~~~ (i = 1,2,3)
    \label{3-28}
\eeq
where $\Om_i$ acts on the Fock space $V_i$ of $V^{\otimes n}$ in (\ref{3-5}).
We assume non-integer conditions on these, $\bra \al_i \ket \not \in {\bf Z}$, as well.
The multivalued function of interest becomes
\beqar
    \Phi  & = &  \Phi_0 \Phi_t
    \nonumber \\
    & = & (-1)^{\frac{1}{\ka}\Om_{12}} (1  - 1/z )^{\frac{1}{\ka}\Om_{23}}
    ( - 1/z )^{\frac{1}{\ka}\Om_{13}} \, t^{-\frac{1}{\ka}\Om_{1}}
    ( 1 - t )^{  - \frac{1}{\ka}\Om_{2}} (  1 - zt )^{  - \frac{1}{\ka}\Om_{3}}
    \label{3-29}
\eeqar
As usually $\Phi$ determines a rank-1 local system
which we denote ${\cal L}_\mathfrak{g}$ to indicate
the algebraic nature of $\Phi$.
The covariant derivative associated with $\Phi$ is expressed as
\beqar
    \nabla_z &=& \d_z - \d_z \log \Phi_0 - \d_z \log \Phi_t
    \nonumber \\
    &=& D_z + \frac{\Om_3 }{\ka} \frac{t}{1-zt}
    \label{3-30}
\eeqar
where $D_z$ is the same as (\ref{3-25c}):
\beq
    D_z \, = \, \d_z - \frac{1}{\ka} \left(
    - \frac{  \Om_{23} + \Om_{13} }{ z } + \frac{ \Om_{13} }{ z-1}
    \right)
    \label{3-31}
\eeq
The covariant derivative (\ref{3-30}) is essentially
the same as the one in (\ref{2-30}) except that
the derivative $\d_z$ is now covariantized as $D_z$.
We should emphasize that, owing to the equivalent relation
(\ref{3-25e}), the action of $D_z$ on a one-form $F(z)dz$ reduces to
$D_z F(z) dz \equiv \d_z F(z) dz$
where $F(z)$ is an arbitrary function (or 0-form) of $z$.
The negative sign in the exponents (\ref{3-28}) is chosen
such that (\ref{3-30}) and (\ref{2-30}) are compatible.

A basis of the cohomology group $H^1 ( X , {\cal L}_\mathfrak{g})$
can be given by a pair $( \varphi_{01}, \varphi_{pq} )$
where $\varphi_{pq} = \{ \varphi_{\infty 0} , \varphi_{1 \frac{1}{z}}, \varphi_{1 \infty},
\varphi_{\frac{1}{z} \infty} , \varphi_{ 0 \frac{1}{z}} \}$;
concrete expressions of these elements are shown in
(\ref{2-26a})-(\ref{2-26c}) and (\ref{2-45a})-(\ref{2-45c}).
Similarly, elements of the homology group $H_1 ( X , {\cal L}_{\mathfrak{g}}^{\vee} )$
can also be obtained from (\ref{2-44}).

For simplicity, we now choose the pair $( \varphi_{01}, \varphi_{\infty 0} )$.
The hypergeometric-type integrals of interest are expressed as
\beqar
    F_{01} (z) &=& \int_\Del \Phi_0 \Phi_t \varphi_{01}
    \label{3-32a} \\
    F_{\infty 0} (z) &=& \int_\Del \Phi_0 \Phi_t \varphi_{\infty 0}
    \label{3-32b}
\eeqar
where $\varphi_{01} = \frac{dt}{t (1 - t)}$, $\varphi_{\infty 0} = \frac{dt }{t}$
and $\Del \in H_1 ( X , {\cal L}_{\mathfrak{g}}^{\vee} )$.
The covariant derivative $D_z$ of $F_{01}$ with respect to $z$ is calculated as
\beq
    D_z \, F_{01} (z) \, = \,
    \int_\Del \Phi_0 \Phi_t \, \nabla_{\! z} \varphi_{01}
    \label{3-33}
\eeq
where $\nabla_{\! z}  = D_z + \frac{\Om_3 }{\ka} \frac{t}{1-zt}$.
The calculation of $\nabla_z \varphi_{01}$ should be executed
under the equivalent condition
\beq
    d \log \Phi_t \, = \,
    - \frac{1}{\ka} \left( \Om_1 \frac{dt }{t} - \Om_2 \frac{dt }{1-t}
    -\Om_3 \frac{z dt}{ 1- zt } \right) \, \equiv \, 0
    \label{3-34}
\eeq
The calculations of $\nabla_{\! z} \varphi_{01}$ and $\nabla_{\! z} \varphi_{\infty 0}$
are therefore exactly the same as those of (\ref{2-32a}) and
(\ref{2-32b}), respectively, except that the exponents are
now replaced by $( a, c-a , -b) \rightarrow ( \frac{1}{\ka} \Om_1 ,
\frac{1}{\ka} \Om_2 , \frac{1}{\ka} \Om_3 )$.
To be concrete, we obtain the following result:
\beq
    D_z
    \left[ \!
      \begin{array}{c}
        F_{01} (z)  \\
        F_{\infty 0} (z) \\
      \end{array}
    \! \right]
    \, = \,
    \int_\Del \Phi_0 \Phi_t \, \nabla_{\! z}
    \left[ \!
      \begin{array}{c}
        \varphi_{01}  \\
        \varphi_{\infty 0}  \\
      \end{array}
    \! \right]
    \, = \,
    \frac{1}{\ka}
    \left(
        \frac{B^{(\infty 0)}_{0}}{z}   + \frac{B^{(\infty 0)}_{1}}{z -1 }
    \right)
    \left[ \!
      \begin{array}{c}
        F_{01} (z)  \\
        F_{\infty 0} (z)  \\
      \end{array}
    \! \right]
    \label{3-35}
\eeq
where
\beq
    B^{(\infty 0)}_{0} =
    \left(
      \begin{array}{cc}
        0 & 0 \\
        \Om_2 & -(\Om_1 + \Om_2 ) \\
      \end{array}
    \right),
    ~~
    B^{(\infty 0)}_{1} =
    \left(
      \begin{array}{cc}
        \Om_2 + \Om_3 & -( \Om_1 + \Om_2 + \Om_3 ) \\
        0 & 0 \\
      \end{array}
    \right)
    \label{3-36}
\eeq
Note that we have used notation $\Om_i$ for
the vacuum expectation value $\bra \Om_i \ket$
in the above expressions.
The differential equation (\ref{3-35}) is first order in the covariant
derivative $D_z$.
Using the results in the previous section, we then find that
$F_{01} (z)$ satisfies
the following {\it covariantized} hypergeometric differential equation
\beq
    \left[
    D_{z}^{2} + \left( \frac{c}{z} + \frac{a+ b+ 1 - c}{z-1} \right) D_z
    + \frac{ab}{z (z-1) }
    \right] F_{01} (z) \,   = \, 0
    \label{3-37}
\eeq
where
\beq
    a = \frac{1}{\ka} \Om_1 \, , ~~
    b = - \frac{1}{\ka} \Om_3 \, , ~~
    c = \frac{1}{\ka} ( \Om_1 + \Om_2 )
    \label{3-38}
\eeq
Notice that, due to the equivalent relation $D_z F_{01} (z) dz \equiv \d_z F_{01} (z) dz$,
the covariantized equation (\ref{3-37}) becomes the ordinary
hypergeometric differential equation (\ref{2-12}) when
evaluated in the integrand over the physical configuration space $\C$ for $n=3$.

An integral representation of the 4-point KZ solution can be constructed as
\beqar
    \Psi ( \infty , 0 ,1, 1/ z ) & = &
    z^{-\frac{1}{\ka} ( \Om_{12} + \Om_{13} + \Om_{23} )} \int_\Del \Phi_0 \Phi_t
    \left[ \!
      \begin{array}{c}
        \varphi_{01}  \\
        \varphi_{\infty 0} \\
      \end{array}
    \! \right]
    \nonumber \\
    &=& z^{-\frac{1}{\ka} ( \Om_{12} + \Om_{13} + \Om_{23} )}
    \left[ \!
      \begin{array}{c}
        F_{01} (z) \\
        F_{\infty 0} (z) \\
      \end{array}
    \! \right]
    \label{3-39}
\eeqar
From this representation we can easily compute $D_z \Psi$ as
\beqar
    D_z \Psi  & = &
     z^{-\frac{1}{\ka} ( \Om_{12} + \Om_{13} + \Om_{23} )}
    \left[ D_z - \frac{1}{\ka}
    \left(
    \frac{ \Om_{12}  }{ z} + \frac{ \Om_{23} }{z -1}
    \right)
    \right]
    \left[ \!
      \begin{array}{c}
        F_{01} (z) \\
        F_{\infty 0} (z) \\
      \end{array}
    \! \right]
    \nonumber \\
    &=&
     z^{-\frac{1}{\ka} ( \Om_{12} + \Om_{13} + \Om_{23} )}
    \frac{1}{\ka}
    \left[
    \left(
        \frac{B^{(\infty 0)}_{0}}{z}   + \frac{B^{(\infty 0)}_{1}}{z -1 }
    \right)
     -
    \left(
    \frac{ \Om_{12}  }{ z} + \frac{ \Om_{23} }{z -1}
    \right)
    \right]
    \left[ \!
      \begin{array}{c}
        F_{01} (z) \\
        F_{\infty 0} (z) \\
      \end{array}
    \! \right]
    \label{3-40}
\eeqar
Thus we confirm that $\Psi$ in (\ref{3-39}) satisfies the KZ equation
$D_z \Psi = 0$ when $\Om_{12}$ and $\Om_{23}$
are represented by the $2 \times 2$ matrices
$B_{0}^{(\infty 0)}$ and $B_{1}^{(\infty 0)}$, respectively.
As mentioned earlier, the basis of the cohomology group $H^1 ( X , {\cal L}_\mathfrak{g})$
can be given by a pair $( \varphi_{01}, \varphi_{pq} )$
where $\varphi_{pq} = \{ \varphi_{\infty 0} , \varphi_{1 \frac{1}{z}}, \varphi_{1 \infty},
\varphi_{\frac{1}{z} \infty} , \varphi_{ 0 \frac{1}{z}} \}$.
Accordingly, we can construct $B_{0}^{(pq)}$, $B_{1}^{(pq)}$
from $A_{0}^{(pq)}$, $A_{1}^{(pq)}$ in the previous section,
with the replacements in (\ref{3-38}).
In general, $\Om_{12}$ and $\Om_{23}$ should be related to these
$B_{0}^{(pq)}$ and $B_{1}^{(pq)}$, respectively.

So far, we argue the 4-point KZ solutions in terms of the
parametrization $(z_0 , z_1 , z_2 , z_3 ) = (\infty , 0,1 , 1/z )$
so that we can utilize the results in the previous section.
Now that the integral representation of the solutions
become clear, we consider the simplest parametrization
$(z_0 , z_1 , z_2 , z_3 ) = (\infty , 0, z, 1 )$ again.
As discussed in (\ref{3-11}), general solutions of the KZ equation
is expressed as $\la \Phi_0$ $( \la \in {\bf C}^\times )$ for
any $n$. Thus the essential part of solving the KZ equation is to find $\la$,
in the present case, in a form of integrals.
From the above analyses we find that for $n=3$ we have two independent
solutions and these can be expressed in terms of the hypergeometric-type
integrals.
The key ingredient of the integral is
the elements of cohomology and homology groups
$H^1 ( X , {\cal L}_{\mathfrak{g}} )$,
$H_1 ( X , {\cal L}_{\mathfrak{g}}^{\vee} )$
where $X = \cp^1 - \{ \infty, 0 , z ,1 \}$ and the
rank-1 local system ${\cal L}_{\mathfrak{g}}$
and its dual ${\cal L}_{\mathfrak{g}}^{\vee}$ are determined by
$\Phi_t$. We now {\it rewrite} $\Phi_t$ as
\beq
    \Phi_t \, = \, t^{- \frac{1}{\ka} \Om_1 } ( t - z )^{- \frac{1}{\ka} \Om_2 }
    (t -1 )^{- \frac{1}{\ka} \Om_3 }
    \label{3-41}
\eeq
The element of the homology group $\Del = H_1 ( X , {\cal L}_{\mathfrak{g}}^{\vee} )$
or the twisted cycle defines the integral path over $X$.
The possible twisted cycles are given by $\Del_{pq}$ where
$(p,q)$ denotes distinct pairs of $\{ \infty , 0 , z , 1 \}$.
The basis of the cohomology group, on the other hand, defines
a one-form to be integrated apart from the factor of the
multivalued function $\Phi = \Phi_0 \Phi_t$.
Owing to the equivalent relation $d \log \Phi_t \equiv 0$,
the basis consists of two elements of $H_1 ( X , {\cal L}_{\mathfrak{g}}^{\vee} )$.
As considered earlier, there are six different choices for the bases.
From (\ref{3-41}) we can easily find the one of these can be given by
$\{ \frac{dt}{t} , \frac{dt }{ t -z } \}$.
The two independent 4-point KZ solutions can then be expressed as
\beq
    \Psi (\infty , 0, z, 1 ) \, = \,
    \int_\Del \Phi_0 \Phi_t
    \left(
      \begin{array}{c}
        \frac{dt}{t} \\
        \frac{dt}{t-z} \\
      \end{array}
    \right)
    \label{3-42}
\eeq
where $\Phi_0 = ( - z )^{\frac{1}{\ka} \Om_{12} } (-1)^{\frac{1}{\ka} \Om_{13}}
(z-1)^{\frac{1}{\ka} \Om_{23}}$ and $\Phi_t $ is given by (\ref{3-41}).
As discussed below (\ref{3-22}), the solution
allows a phase factor
arising from $(z_3 - z_1 )^{\frac{1}{\ka}( \Om_{12} + \Om_{13} + \Om_{23})}$.

To summarize, the 4-point KZ solutions can be given by
the following generalized hypergeometric functions on $Gr ( 4, 2)$:
\beqar
    F_j ( z_0 , z_1 , z_2 , z_3 )  & = &
    \int_\Del \Phi_0 \Phi_t \, \varphi_j
    \label{3-43a} \\
    \Phi_0 &=& \prod_{1 \le i < j \le 3} ( z_i - z_j )^{\frac{1}{\ka} \Om_{ij}}
    \label{3-43b} \\
    \Phi_t &=& \prod_{i = 1}^{3} ( t - z_i )^{- \frac{1}{\ka} \Om_i }
    \label{3-43c}
\eeqar
where $\Del \in H_1 ( X , {\cal L}_{\mathfrak{g}}^{\vee} )$ and $\varphi_j \in
H^1 ( X , {\cal L}_{\mathfrak{g}} )$. From our study on the
generalized hypergeometric functions on $Gr (4,2)$, we find
the elements $\varphi_j$ are given by the following set
\beq
    \varphi_j \, = \, \left\{
    \frac{dt }{t - z_1} ,\frac{dt }{t - z_2} , \frac{dt }{t - z_3},
    \frac{(z_1  - z_2 ) dt}{(t-z_1 )(t - z_2 )},
     \frac{(z_1  - z_3 )dt}{(t-z_1 )(t - z_3 )},
    \frac{(z_2  - z_3 ) dt}{(t-z_2 )(t - z_3 )}
    \right\}
    \label{3-44}
\eeq
The 4-point KZ solutions have two independent solutions. These are obtained
by choosing two elements from the above. The simplest choice may be
$\{ \frac{dt}{t - z_1 } , \frac{dt }{ t - z_2 } \}$. This corresponds
to the solutions in (\ref{3-42}).

\vskip 0.5cm \noindent
\underline{Relation to $(n+1)$-point KZ solutions}

At the present stage, it is straightforward to generalize the
above results to $(n+1)$-point solutions of the KZ equation.
These are obtained as generalized hypergeometric functions
on $Gr ( 2 , n+1)$. Following the representation (\ref{3-43a})-(\ref{3-44}),
we can write down the $(n+1)$-point KZ solutions as
\beqar
    F_j ( z_0 , z_1 , z_2 , \cdots , z_n )  & = &
    \int_\Del \Phi_0 \Phi_t \, \varphi_j
    \label{3-45a} \\
    \Phi_0 &=& \prod_{1 \le i < j \le n} ( z_i - z_j )^{\frac{1}{\ka} \Om_{ij}}
    \label{3-45b} \\
    \Phi_t &=& \prod_{i = 1}^{n} ( t - z_i )^{- \frac{1}{\ka} \Om_i }
    \label{3-45c}
\eeqar
where $\Phi_t$ is now defined on $X = \cp^1 - \{ z_0 , z_1 , \cdots , z_n \}$.
The construction of $F_j$ is essentially the same as the one considered in (\ref{2-9}).
From $\Phi_t$ we can determine the homology group $H_1 ( X , {\cal L}_{\mathfrak{g}}^{\vee} )$
and the cohomology group $H^1 ( X , {\cal L}_{\mathfrak{g}} )$.
The twisted cycle $\Del$ and the one-form $\varphi_j$ are elements of these, respectively, {\it i.e.},
$\Del \in H_1 ( X , {\cal L}_{\mathfrak{g}}^{\vee} )$ and $\varphi_j \in
H^1 ( X , {\cal L}_{\mathfrak{g}} )$.
The number of elements for the basis of the cohomology group is
$n-1$ and such a basis can be chosen as
\beq
    \varphi_j \, = \, d \log \frac{ t - z_{j+1} }{ t - z_j } ~~~~~~~~~
    ( 1 \le j \le n-1 )
    \label{3-46}
\eeq
There are $n-1$ solutions and these correspond to $n-1$ independent
solutions of the $(n+1)$-point KZ equation.
Namely, we can express the $(n+1)$-point KZ solution as
\beq
    \Psi \, = \,
    \left(
      \begin{array}{c}
        F_1 \\
        F_2 \\
        \vdots \\
        F_{n-1} \\
      \end{array}
    \right)
    \label{3-47}
\eeq

As in the $n = 3$ case, it is known that this $\Psi$ satisfy the
differential equation $d \Psi = B \Psi$ where $B$ is called the Gauss-Manin
connection and represented by an $(n+1) \times (n+1)$ matrix.
This Gauss-Manin connection  is associated to the
definition of the generalized hypergeometric functions on $Gr ( 2 , n+1)$.
In general, such a Gauss-Manin connection can be constructed
in association with the generalized hypergeometric functions on $Gr ( k + 1 , n+1)$,
where the dimension of $B$ is given by $
\left( \!\!
  \begin{array}{c}
    n-1 \\
    k \\
  \end{array} \!\!
\right)
$.
The study of the Gauss-Manin connection is beyond the scope of this note.
Interested readers are advised to see mathematical literature, {\it e.g.},
Section 3.8 in \cite{Aomoto:1994bk}.
In this context, the KZ connection (\ref{3-13}) can be interpreted as
the Gauss-Manin connection for the $Gr ( 2 , n+1)$-type generalized hypergeometric functions.

So far, we consider the case of $k = 1$. This is natural because it clarifies
the relation between Gauss' hypergeometric function
and the 4-point KZ solutions.
For general $(n+1)$-point KZ solutions, however, there are no particular
reasons to choose $k = 1$ except it leads to the simplest hypergeometric integrals.
In principle, we can consider $k \ge 2$ cases and relate the KZ solutions to
generalized hypergeometric functions on $Gr ( k + 1, n+1 )$.
In fact, there is a remarkable result or a theorem by
Schechtman and Varchenko \cite{Schechtman:1989,Schechtman:1990zc}
that, with  $\mathfrak{g}$ being the $SL(2, {\bf C})$ algebra,
the KZ solutions can be expressed by the following hypergeometric-type integral
\beqar
    F_{ J }   &= &
    \int_\Del
    \prod_{1 \le i < j \le n} ( z_i - z_j )^{\frac{1}{\ka} \Om_{ij} }
    \prod_{j = 1}^{n} \prod_{s=1}^{k} ( t_s - z_j )^{\frac{1}{\ka} \Om_j }
    \nonumber \\
    && ~\times ~
    \prod_{1 \le r < s \le k} ( t_r - t_s )^{\frac{1}{\ka} }
    \,  R_J ( t, z) \,  dt_1 \wedge \cdots \wedge d t_k
    \label{3-48}
\eeqar
In this expression a multivalued function relevant to $\Phi_t$ in (\ref{3-45c})
is given by
\beq
    \widetilde{\Phi}_t \, = \,
    \prod_{j = 1}^{n} \prod_{s=1}^{k} ( t_s - z_j )^{\frac{1}{\ka} \Om_j }
    \, = \, \prod_{j = 1}^{n} {\widetilde{l}_{j} (t)}^{\frac{1}{\ka} \Om_j }
    \label{3-49}
\eeq
where
\beq
    \widetilde{l}_j (t) \,  = \, (t_1 - z_j )(t_2 - z_j ) \cdots (t_k - z_j )
    \label{3-50}
\eeq
This multivalued function $\widetilde{\Phi}_t$ is defined on the space
\beq
    \widetilde{X} \, = \, {\bf C}^k - \bigcup_{j= 1}^{n} \widetilde{{\cal H}}_j
    \label{3-51}
\eeq
where
\beq
    \widetilde{{\cal H}}_j \, = \,  \{ t \in {\bf C}^k \, ; ~ \widetilde{l}_j (t) = 0 \}
    \label{3-52}
\eeq
Namely, $\widetilde{X}$ represents a coordinate on
${\bf C}^k$, eliminating $n$ distinct points $(t_1 , t_2 , \cdots , t_k) =
( z_{j}, z_j , \cdots , z_j) $ for $1 \le j \le n$.
This space is analogous to the one in (\ref{1-38}) except that we have
a different $\widetilde{l}_j (t)$ here.
From $\widetilde{\Phi}_t$ we can then define the $k$-th homology and cohomology groups,
$H_k ( \widetilde{X}, \widetilde{\cal L}_{\mathfrak{g}}^{\vee})$
and $H^k ( \widetilde{X}, \widetilde{\cal L}_{\mathfrak{g}} )$.
In the integral (\ref{3-48}), $\Del$ denotes an element of
$H_k ( \widetilde{X}, \widetilde{\cal L}_{\mathfrak{g}}^{\vee})$
or a twisted $k$-cycle.

According to Schechtman and Varchenko \cite{Schechtman:1989,Schechtman:1990zc},
$R_J ( t ,z )$ in (\ref{3-48}) is a rational function of $t_s$ and $z_j$ and is
expressed as follows (see also recent reviews \cite{Kohno:2012a, Kohno:2012b}).
Let $J$ be a set of $n$ non-negative integers $J = ( j_1 , j_2 , \cdots , j_n )$
under the condition $|J| = j_1 + j_2 + \cdots + j_n = k$.
Note that there is no particular maximum limit for $k$ (such as $k < n$)
in the original derivation \cite{Schechtman:1989,Schechtman:1990zc}.
(In an alternative derivation with a free field OPE method
\cite{Awata:1991az}, $k$ corresponds to the number of insertions
in an $(n+1)$-point correlators.)
Thus, for $j_i \in {\bf Z}_{\ge 0}$ ($i=1, 2, \cdots , n$), we have
$\left( \!\!
  \begin{array}{c}
    n + k -1 \\
    n -1  \\
  \end{array} \!\!
\right)$
different elements in $J$. This corresponds to the number of
possible $n$-partitions of integer $k$, allowing an empty set.
To be concrete, for each choice of $J = ( j_1 , j_2 , \cdots , j_n )$,
we can define an $n$-partition of
the sequence of $k$ integers $(s_1 , s_2 , \cdots , s_k )$
such that
\beq
    (s_1 , s_2 , \cdots , s_k ) = \Bigl( \underbrace{1, \cdots , 1}_{j_1},
    \underbrace{2, \cdots, 2}_{j_2} , \cdots ,\underbrace{n, \cdots, n}_{j_n}
    \Bigr)
    \label{3-53}
\eeq
Accordingly, we can define a rational function
\beq
    S_J ( z_1 , \cdots , z_n , t_1 , \cdots , t_k )
    \, = \,
    \frac{1}{ ( t_1  - z_{s_1} ) ( t_2 - z_{s_2} ) \cdots ( t_k - z_{s_k} ) }
    \label{3-54}
\eeq
The rational function $R_J ( t ,z)$ is then defined as
\beq
    R_J (t ,z ) \, = \, \frac{1}{ j_1 ! j_2 ! \cdots j_n ! }
    \sum_{\si \in \S_k } S_J ( z_1 , \cdots , z_n, t_{\si_1} , \cdots , t_{\si_k} )
    \label{3-55}
\eeq
where the summation of $\S_k$ is taken over the permutations
$\si = \left(
         \begin{array}{c}
           1 ~ 2 ~ \cdots \, k \\
           \si_1 \si_2 \cdots \si_k \\
         \end{array}
       \right)$.
For simple cases, $R_J ( t, z)$ are written down as
\beqar
    && R_{(1,0,\cdots, 0)} ( z, t) = \frac{1}{t_1 - z_1 }\, , ~~
    R_{(2,0,\cdots, 0)} ( z, t) = \frac{1}{(t_1 - z_1 )(t_2 - z_1 ) }\, ,
    \nonumber \\
    && R_{(1,1, 0, \cdots, 0)} ( z, t) = \frac{1}{( t_1 - z_1 )(t_2 - z_2 ) }
    + \frac{1}{( t_2 - z_1 )(t_1 - z_2 ) }
    \label{3-56}
\eeqar
Schechtman and Varchenko show that the $(n+1)$-point KZ solutions
can be given by
\beq
    \sum_{|J| = k } F_J
    \label{3-57}
\eeq
for arbitrary $k$ and $F_J$ is defined in (\ref{3-48}).

In what follows we show that
$R_J ( t , z) dt_1 \wedge \cdots \wedge dt_k$ can be interpreted
as an element of the $k$-th cohomology group
$H^k ( \widetilde{X}, \widetilde{\cal L}_{\mathfrak{g}} )$
for $k < n $ and  $j_i \in \{ 0 , 1 \}$ ($i=1, 2, \cdots , n$).
The condition $k < n$ is necessary to
relate the solutions to generalized hypergeometric functions
on $Gr ( k+1, n+1)$.
The other condition $j_i \in \{ 0 , 1 \}$ ($i=1, 2, \cdots , n$) arises from the
fact that the number of the basis for
$H^k ( \widetilde{X}, \widetilde{\cal L}_{\mathfrak{g}} )$
is given by
$
\left( \!\!
  \begin{array}{c}
    n-1 \\
    k \\
  \end{array} \!\!
\right)
$ as mentioned in (\ref{1-40}) and below.
Under these conditions the label $J$ can be replaced by a set of $k$ integers
$\{ j_1 , j_2 , \cdots , j_k \}$, satisfying
$ 1 \le j_1 < j_2 < \cdots < j_k \le n-1$.
The basis of $H^k ( \widetilde{X}, \widetilde{\cal L}_{\mathfrak{g}} )$
is then given by
\beqar
    \varphi_{j_1 , j_2 , \cdots , j_k }
    & = &
    d \log \widetilde{l}_{j_1} \wedge d \log \widetilde{l}_{j_2}
    \wedge \cdots \wedge d \log \widetilde{l}_{j_k}
    \nonumber \\
    &=&
    \left(
    \frac{d t_1}{ t_1  - z_{j_1} } + \cdots + \frac{ dt_k }{ t_k - z_{j_1} }
    \right) \wedge
    \left(
    \frac{d t_1}{ t_1  - z_{j_2} } + \cdots + \frac{ dt_k }{ t_k - z_{j_2} }
    \right)
    \nonumber \\
    &&
    ~~ \wedge
    \cdots \wedge
    \left(
     \frac{d t_1}{ t_1  - z_{j_k} } + \cdots + \frac{ dt_k }{ t_k - z_{j_k} }
    \right)
    \nonumber \\
    &=&
    \sum_{\si \in \S_k } \frac{ dt_1 \wedge dt_2 \wedge \cdots \wedge d t_k}{
    ( t_1 - z_{j_{\si_1}} ) ( t_2 - z_{j_{\si_2}} ) \cdots ( t_k - z_{j_{\si_k}} )
    }
    \label{3-58}
\eeqar
In comparison with (\ref{3-55}), we find that
$\varphi_{j_1 , j_2 , \cdots , j_k }$ are equivalent to
$R_I ( t, z)$ where the elements of $I  = ( i_1 , i_2 , \cdots , i_n )$
are set to $i_l = 1$ for $l = j_1 , j_2 , \cdots , j_k $ and $i_l = 0$ otherwise.
This illustrates a direct relation of the $(n+1)$-point KZ solutions
to the hypergeometric integrals of a form
\beqar
    F_{j_1 j_2 \cdots j_k } &=&
    \int_\Del  \widetilde{\Phi}_0 \widetilde{\Phi}_t \, \varphi_{j_1 j_2 \cdots j_k}
    \label{3-59a}\\
    \widetilde{\Phi}_0 &=&
    \prod_{1 \le i < j \le n} ( z_i - z_j )^{\frac{1}{\ka} \Om_{ij} }
    \prod_{1 \le r < s \le k} ( t_r - t_s )^{\frac{1}{\ka} }
    \label{3-59b} \\
    \widetilde{\Phi}_t & = &
    \prod_{j = 1}^{n} {\widetilde{l}_{j} (t)}^{\frac{1}{\ka} \Om_i }
    \label{3-59c}
\eeqar
where $\widetilde{l}_{j} (t)$ is defined by (\ref{3-50}).
Notice that the basis $\varphi_{j_1 j_2 \cdots j_k}$ in (\ref{3-58}) is not
the unique choice. We can choose different bases which lead
to alternative parametrization of the solutions $F_{j_1 j_2 \cdots j_k }$
For example, using the result in (\ref{1-41}),
we can also express the basis of $H^k ( \widetilde{X}, \widetilde{\cal L}_{\mathfrak{g}} )$ as
\beqar
    \varphi_{ j_1 , j_2 , \cdots , j_k }
    &=&
    d \log \frac{ \widetilde{l}_{j_1 + 1} }{ \widetilde{l}_{j_1 } }
    \wedge
    d \log \frac{ \widetilde{l}_{j_2 + 1} }{ \widetilde{l}_{j_2 } }
    \wedge
    \cdots
    \wedge
    d \log \frac{ \widetilde{l}_{j_k + 1} }{ \widetilde{l}_{j_k } }
    \nonumber \\
    &=&
    ( z_{j_1 +1} - z_{j_1} )( z_{j_2 +1} - z_{j_2} )
    \cdots ( z_{j_k +1} - z_{j_k} )
    \nonumber \\
    && \hspace{-0.6cm}
    \times
    \sum_{\si \in \S_k }
    \frac{ dt_1 \wedge dt_2 \wedge \cdots \wedge dt_k }{
    ( t_1 - z_{j_{\si_1} + 1} )( t_1 - z_{j_{\si_1}} ) ( t_2 - z_{j_{\si_2}+1} ) ( t_2 - z_{j_{\si_2}} )
    \cdots ( t_k - z_{j_{\si_k}+1} ) ( t_k - z_{j_{\si_k}} )
    }
    \nonumber \\
    \label{3-60}
\eeqar
In general, the basis can be chosen as
$\varphi_{ j_1 , j_2 , \cdots , j_k } =
    d \log \frac{ \widetilde{l}_{j_1 + a} }{ \widetilde{l}_{j_1 } }
    \wedge
    \cdots
    \wedge
    d \log \frac{ \widetilde{l}_{j_k + a} }{ \widetilde{l}_{j_k } }
$ where $a \equiv 1,2, \cdots ,n-1$ (mod $n$).

Lastly, we notice that $\widetilde{l}_j (t)$ in (\ref{3-50}) is linear in terms of the elements of
$t = (t_1 , t_2 , \cdots , t_k )$. This enable us to determine the rank-1 local systems
$\widetilde{\cal L}_{\mathfrak{g}}$, $\widetilde{\cal L}_{\mathfrak{g}}^{\vee}$
in association to $\widetilde{\Phi}_t$ and make it straightforward to
construct the hypergeometric integrals (\ref{3-59a}).
However, if we consider
\beq
    \prod_{1 \le r < s \le k} ( t_r - t_s )^{\frac{1}{\ka} }
    \, \widetilde{\Phi}_t
    \label{3-62}
\eeq
rather than $\widetilde{\Phi}_t$, as
a multivalued function of interests, we can not
properly determine rank-1 local systems out of it since
it can not be factorized into functions linear in $t$.
In order to circumvent this issue,
one may regard the above multivalued function as a
function on ${\bf C}^{n+k}$. But this leads to mixture
of variables in $t_s$ and $z_i$ and a resulting
hypergeometric integral may be regarded as that on
$Gr ( 1, n + k +1)$.
Thus it is not appropriate to think of (\ref{3-62})
as a multivalued function of interest when we interpret
$F_{j_1 j_2 \cdots j_k}$ as hypergeometric functions.

One may still wonder why $\widetilde{\Phi}_t$ instead of (\ref{3-62}) should be
extracted as the defining multivalued function.
The author do not have a satisfying answer to it;
this could be an ambiguity in the construction of
$F_{j_1 j_2 \cdots j_k}$. This issue  arises from the fact
that $F_{j_1 j_2 \cdots j_k}$ is not {\it exactly} defined
as generalized hypergeometric function on $Gr ( k+1 ,n+1)$.
As discussed in the beginning of this section, the configuration
space of the KZ solutions $\Psi ( z_0 , z_1 , z_2 ,\cdots ,z_n)$ is equivalent
to that of generalized hypergeometric functions on $Gr ( 2 , n+1)$,
which is represented by $n+1$ distinct points in $\cp^1$.
If we relate the $(n+1)$-point KZ solutions
to generalized hypergeometric functions on $Gr(k+1, n+1)$ in
a rigorous manner, we need to expand the configuration space
such that it is represented by $n+1$ distinct points in $\cp^k$
but this brings about ambiguities with the actual/physical
configuration space mentioned above.

In conclusion, we can express solutions of the KZ equation
in terms of hypergeometric-type integrals.
The $(n+1)$-point KZ solutions in general can be
represented by generalized hypergeometric functions
on $Gr( 2 , n+1)$ as shown in (\ref{3-45a})-(\ref{3-45c}).
We can generalize this expression to represent the $(n+1)$-point KZ solution
as hypergeometric-type integrals on $Gr (k+1, n+1)$, as shown in (\ref{3-59a})-(\ref{3-59c}),
but there exists a subtle ambiguity in rigorous construction of the integrals
for $k \ge 2$.

\section{Holonomy operators of KZ connections}

In the present section we review the construction of holonomy
operators of the Knizhnik-Zamolodchikov (KZ) connection,
following \cite{Kohno:2002bk,Abe:2009kn}, and consider the holonomy operators
in relation to cohomology and homology of the physical configuration
space $\C$ in (\ref{3-17}).
We first reconsider the KZ equation.
We rewrite the KZ equation (\ref{3-1}) as a differential
form (\ref{3-12}), $D \Psi = ( d  - \Om ) \Psi= 0$,
where the KZ connection $\Om$ is defined as
\beqar
    \Om &=&  \frac{1}{\kappa} \sum_{1 \le i < j \le n} \Om_{ij} \, \om_{ij}
    \label{6-1} \\
    \om_{ij} &=& d \log (z_i - z_j) = \frac{ d z_i - d z_j}{z_i - z_j}
    \label{6-2}
\eeqar
In doing so, we implicitly use the condition $\Om_{ij} = \Om_{ji}$.
The KZ equation is originally derived from the application
of a Ward identity to the current correlators.
Action of the operator $\Om_{ij}$ on the Hilbert space
$V^{\otimes n} = V_1 \otimes V_2 \otimes \cdots \otimes V_n$
is then defined as
\beq
    \sum_{\mu} 1 \otimes \cdots \otimes 1 \otimes \rho_i (I_{\mu})
    \otimes 1 \otimes \cdots \otimes 1 \otimes \rho_j (I_{\mu}) \otimes 1 \otimes \cdots
    \otimes 1
    \label{6-3}
\eeq
where $I_\mu$ ($\mu = 1,2, \cdots , \dim \mathfrak{g}$) are elements of the
Lie algebra $\mathfrak{g}$
and $\rho$ denotes its representation.
Thus, by definition, $\Om_{ij}$ satisfies $\Om_{ij} = \Om_{ji}$.

For example, in the conventional choice of $\mathfrak{g}$ being the $SL(2, {\bf C})$ algebra
$\Om_{ij}$ can be defined as
\beq
    \Om_{ij} = a_{i}^{(+)} \otimes a_{j}^{(-)} + a_{i}^{(-)} \otimes a_{j}^{(+)}
    + 2 a_{i}^{(0)} \otimes a_{j}^{(0)} 
    \label{6-4}
\eeq
where the operators $a_{i}^{(\pm , 0)}$ act on the $i$-th Fock space $V_i$
and forms the $SL(2, {\bf C})$ algebra:
\beq
    [ a_{i}^{(+)}, a_{j}^{(-)}] = 2 a_{i}^{(0)} \, \del_{ij}  \, , ~~
    [ a_{i}^{(0)}, a_{j}^{(+)}] = a_{i}^{(+)} \, \del_{ij} \, , ~~
    [ a_{i}^{(0)}, a_{j}^{(-)}] = - a_{i}^{(-)} \, \del_{ij}
    \label{6-5}
\eeq
where $\del_{ij}$ denotes Kronecker's delta.
Note that in the case of $i = j$, $\Om_{ii}$ becomes the quadratic Casimir of $SL(2, {\bf C})$ algebra
which acts on $V_i$. This defines the operator $\Om_i$ that we have introduced in (\ref{3-28}).

The resultant KZ solutions in an integral form
show that the solutions can be described in terms
of $\Om_{ij}$ where $1 \le i < j \le n$. This is a natural consequence
of our setting that the physical configuration space
of the KZ solution $\C$ can be represented by ordered
distinct $n+1$ points in $\cp^1$.
We have already considered such a space $\C = X_n / \S_n$ in (\ref{3-17}) and see
that the monodromy representation of the KZ equation
is given by the braid group $\B_n = \Pi_1 ( \C )$.
From these perspectives we can discard the operators $\Om_{ji}$ $(i<j)$
and begin with $D \Psi = ( d  - \Om ) \Psi= 0$ as the defining KZ equation.
{\it The study of a KZ system
can then be attributed to the classification of the KZ connections
which satisfy the infinitesimal braid relations (\ref{3-6a}) and (\ref{3-6b}).}
As shown in (\ref{3-16}), such a KZ connection becomes a flat connection, $D \Om= 0$.
Thus, a general solution of the KZ equation can
be given by a holonomy of $\Om$. In the language of
gauge theory the holonomy is given by a Wilson loop
operator of a gauge field in question. According to Kohno \cite{Kohno:2002bk},
the holonomy of $\Om$ provides a general linear representation
of the braid group on the Hilbert space $V^{\otimes n}$.

\vskip 0.5cm \noindent
\underline{Holonomy operators of the KZ connections: a review}

The holonomy of $\Om$ can be defined as \cite{Kohno:2002bk}:
\beq
    \Theta_{\ga} \, = \, 1 + \sum_{r \ge 1} \oint_\ga
    \underbrace{ \Om \wedge \Om \wedge \cdots \wedge \Om}_{r}
    \label{6-6}
\eeq
where $\ga$ represents a closed path on $\C_r = X_r / \S_r$
where $X_r$ is defined in (\ref{3-3}).
In the following we shall denote the physical configuration space,
with the the number of dimensions being explicit.
Since the integrand in (\ref{6-6}) is an $r$-form, the corresponding
integral is taken over the $r$-dimensional complex space $\C_r$.
Formally, the above integral can be evaluated as an iterated integral
of K.~-T.~Chen \cite{Chen:1977oja}.
Let the path $\ga$ in $\C_r$ be represented by
\beq
    \ga(t) = ( z_1 ( t  ) , z_2 (t ) , \cdots , z_r (t) ) ~~~~~~~~~
    0 \le t \le 1
    \label{6-7}
\eeq
Denoting the pull-back $\ga^* \om_{ij}$ as
\beq
    \ga^* \om_{ij} \, = \,    \om_{ij} (t) \, = \,
    \frac{ d z_i (t) - d z_j (t) }{ z_i (t) - z_j (t) } \, ,
    \label{6-8}
\eeq
we can explicitly express $\Theta_\ga$ as an iterated integral
\beq
    \Theta_\ga \, = \,
    \sum_{r \ge 0 } \frac{1}{\ka^r}
    \int_{ 0 \le t_1 \le t_2 \le \cdots \le t_r \le 1}
    \sum_{ (i<j) } \Om_{i_1 j_1} \Om_{i_2 j_2} \cdots \Om_{i_r j_r}
    \bigwedge_{l=1}^{r} \om_{i_l j_l} (t_l)
    \label{6-9}
\eeq
where $(i < j)$ means that the set of indices $(i_1, j_1, \cdots , i_r , j_r)$
are ordered such that $1 \le i_l < j_l \le r$ for $l= 1,2, \cdots, r$.

Let the initial point of $\ga (t) $ be $z_i (0) = z_i$ for $i=1,2,\cdots , r$
and the final point be $(z_1 (1), z_j (1) )= (z_1, z_{\si_j})$ for $j=2,3, \cdots, r$
and $\si=
\left( \!\!
\begin{array}{c}
  2 \, ~ 3 ~ \cdots ~ r \\
  \si_2 \si_3 \cdots \si_r \\
\end{array} \!\!
\right)$.
Since $\C_r = X_r / \S_r$ is permutation invariant,
the initial and final points are identical and
we can naturally interpret $\ga$ as a closed path on $\C_r$.
By definition $\ga$ represents an element of
the braid group:
\beq
    \ga \, \in  \, \Pi_1 ( \C_r  ) = \B_r
    \label{6-10}
\eeq
To make $\Theta_\ga$ permutation invariant explicitly, we now
{\it redefine} the holonomy operator of $\Om$ as an analog
of the Wilson loop operator in gauge theory \cite{Abe:2009kn}:
\beq
    \Theta_{\ga}  = \Tr_{ \ga} \, \Path \exp \left[
    \sum_{r \ge 2} \oint_{\ga} \underbrace{ \Om \wedge \Om \wedge \cdots \wedge \Om}_{r}
    \right]
    \label{6-11}
\eeq
The meanings of the symbol $\Path$ and the trace $\Tr_{\ga}$ are clarified below.
As in (\ref{6-9}), the exponent in (\ref{6-10}) can be expanded as
\beq
    \oint_{\ga} \underbrace{\Om \wedge \cdots \wedge \Om}_{r}
    \, = \,
    \frac{1}{\ka^r} \oint_{\ga}
    \sum_{ (i<j) } \Om_{i_1 j_1} \Om_{i_2 j_2} \cdots \Om_{i_r j_r}
    \, \om_{i_1 j_1} \wedge \om_{i_2 j_2} \wedge \cdots \wedge \om_{i_r j_r}
    \label{6-12}
\eeq
Action of the symbol $\Path$ on the above integral imposes the ordering conditions
$1 \le i_1 < i_2 < \cdots < i_r \le r$ and $2 \le j_1 < j_2 < \cdots < j_r \le r+1$,
with $r+1 \equiv 1$ (mod $r$). Thus we have
\beq
    \Path \oint_{\ga} \underbrace{\Om \wedge \cdots \wedge \Om}_{r}
    \, = \,
    \frac{1}{\ka^r} \oint_{\ga}  \Om_{1 2} \Om_{2 3} \cdots \Om_{r 1}
    \, \om_{12} \wedge \om_{23} \wedge \cdots \wedge \om_{r 1}
    \label{6-13}
\eeq
The trace $\Tr_{\ga}$ in (\ref{6-11})
is carried out by summing over permutations of the indices:
\beq
    \Tr_{\ga} \Path  \oint_{\ga}
    \underbrace{\Om \wedge \cdots \wedge \Om}_{r}
    \, = \,
    \sum_{\si \in \S_{r-1}} \frac{1}{\ka^r} \oint_{\ga}  \Om_{1 \si_2} \Om_{\si_2 \si_3} \cdots \Om_{\si_r 1}
    \, \om_{1 \si_2} \wedge \om_{\si_2 \si_3} \wedge \cdots \wedge \om_{\si_r 1}
    \, := \, I_r
    \label{6-14}
\eeq
This can be interpreted as a trace over generators of the braid group $\B_r$
and is called a braid trace.

\vskip 0.5cm \noindent
\underline{Homology and cohomology interpretations of the integral $I_r$}

The holonomy operator (\ref{6-11}) is essentially calculated by the above integral $I_r$.
We now consider $I_r$ in terms of cohomology and homology of $\C_r$.
The cohomology part is relatively straightforward.
Since the KZ connection $\Om$ is a flat connection $D \Om = 0$, it is
a closed form with respect to the covariant derivative $D = d - \Om$.
But, by definition, it is not an exact form, that is, $\Om \ne D f $ for
any function $f$ of $(z_1 , \cdots , z_r )$.
Thus it is an element of the cohomology group $H^1 ( \C_r  )$ whose
coefficients are given by $\mathfrak{g} \otimes \mathfrak{g}$,
\beq
    \Om \, \in \, H^1 ( \C_r  , \mathfrak{g} \otimes \mathfrak{g} )
    \label{6-15}
\eeq
To generalize, the integrand of $I_r$ can be considered as
\beq
    \underbrace{ \Om \wedge  \cdots \wedge \Om}_{r}
    \, \in \, H^r ( \C_r  , \mathfrak{g}^{\otimes r} )
    \label{6-16}
\eeq
where $\mathfrak{g}^{\otimes r}$ denotes
a set of operators acting on the Hilbert space $V^{\otimes r}$.

Since $\ga$ is defined as a closed path in $\C_r$, the algebraic
coefficients $\Om_{i_1 j_1}  \cdots \Om_{i_r j_r}$ can be extracted
out of the integrand.
In order to make sense of $\Theta_\ga$ as an integral,
we need to regard $\ga$ as an element of
the $r$-th homology group $H_r ( \C_r , {\bf R} )$, with the coefficients
being real number:
\beq
    \ga \, \in \, H_r ( \C_r , {\bf R} )
    \label{6-17}
\eeq
The element $\ga$ can be considered as a path in $\C_r$
connecting $r$ hyperplanes $\H_{ij}$ defined in (\ref{3-4}).
Since the KZ equation has branch points at $\H_{ij}$ $( i < j )$,
(\ref{6-17}) is in accord with a general concept of
the homology group by use of boundary operators as
discussed in (\ref{1-15})-(\ref{1-17}).
Note that the homology and cohomology considered here are
not the twisted ones as before.
Thus we can not directly relate $I_r$ to the bilinear construction of
hypergeometric-type integrals in (\ref{1-18}) and (\ref{1-19}).
In the present note, however, we have been equipped with
a certain level of understanding of the hypergeometric integrals
in terms of homology and cohomology. Namely, we learned that a
(co)homology interpretation provides a systematic treatments of analytic continuation
or monodromy representation of the solutions to a differential equation of interests.
In what follows, we further consider these aspects of the integral $I_r$.

Let $L \C_r$ be a loop space in $\C_r$. It is known that
the fundamental homotopy group of $\C_r$ is isomorphic to the
0-dimensional homology group of $L \C_r$ \cite{Kohno:2009bk}:
\beq
    \Pi_1 ( \C_r ) \cong H_0 ( L \C_r )
    \label{6-18}
\eeq
With the result (\ref{6-10}), we then find that $\ga$ can also be
an element of $H_0 ( L \C_r )$, with the coefficients being the real number:
\beq
    \ga \, \in \, H_0 ( L \C_r , {\bf R} )
    \label{6-19}
\eeq
This suggests that the integrand of $I_r$ can and should be interpreted
as an element of the 0-dimensional cohomology group of
the loop space $L \C_r$. We can then naturally {\it assume}
\beq
    \underbrace{ \Om \wedge  \cdots \wedge \Om}_{r}
    \, \in \, H^0 ( L \C_r  , \mathfrak{g}^{\otimes r} )
    \label{6-20}
\eeq
In the following, we briefly argue that this assumption is a favorable one.
Remember that the basis of $H^{r-1} ( \widetilde{X}, \widetilde{\cal L}_{\mathfrak{g}} )$
considered in (\ref{3-58}), with $n=r$ and $k = r-1$, can be expressed as
\beq
    \varphi_{2 , 3 , \cdots , r } \, = \,
    \sum_{\si \in \S_{r-1} } \frac{ dt_2 \wedge dt_3 \wedge \cdots \wedge d t_{r}}{
    ( t_2 - z_{\si_2} ) ( t_3 - z_{\si_3} ) \cdots ( t_r - z_{\si_r} )
    }
    \label{6-21}
\eeq
Adding the parameters $(t_1, z_1 )$, we can {\it redefine} the above basis as
\beq
    \varphi_{1, 2 , 3 , \cdots , r } \, = \,
    \sum_{\si \in \S_{r-1} } \frac{
     dt_1 \wedge dt_2 \wedge dt_3 \wedge  \cdots \wedge d t_{r}
    }{
    ( z_1 - t_1 ) ( z_{\si_2} -t_2  ) ( z_{\si_3} - t_3   ) \cdots ( z_{\si_r} - t_r )
    }
    \label{6-22}
\eeq
By definition (see {\it e.g.}, (\ref{1-37b})),
the dimension-0 cohomology means there are no extra
parameters except $\{ z_1 , z_2 , \cdots , z_r \} \in L \C_r$.
The basis of the dimension-0 cohomology group  $H^0 ( L \C_r  )$
can then be obtained by identifying $\{ t_1 , t_2 , \cdots , t_r \}$
with $\{ z_1 , z_2 , \cdots , z_r \}$ in the above.
To avoid divergence, we need to require $t_i \ne z_i$ in
the denominator; since $z_i$'s are permutation invariant we can
choose $t_i$'s arbitrarily such that $\varphi_{1,2,\cdots, r}$ becomes finite.
One of the simplest nontrivial results
can be obtained by setting $t_i = z_{\si_{i+1}}$ in
the denominator of (\ref{6-22}), with $t_r = z_{\si_{r+1}} = z_1$.
We then have
\beq
    \left.
    \varphi_{1,2,\cdots ,r}
    \right|_{t=z}
    \, = \,
    \sum_{\si \in \S_{r-1} }  \frac{
     dz_1 \wedge dz_{2}  \wedge  \cdots \wedge d z_r
    }{
    ( z_1 - z_{\si_2} ) ( z_{\si_2} -  z_{\si_3} ) ( z_{\si_3} - z_{\si_4}  ) \cdots ( z_{\si_r} - z_1 )
    }
    \label{6-23}
\eeq
Apart from the algebraic part involving $\Om_{ij}$, the above factor
is identical to the integral $I_r$ at the level of integrand.
Thus, as far as the integrand is concerned,
the fact that the basis of
$H^0 ( L \C_r  )$ is given by (\ref{6-23}) leads to
a favorable confirmation of the assertion (\ref{6-20}).

As discussed earlier, we need to determine a multivalued
function to make a (co)homology interpretation of the integral $I_r$.
The above analyses show that a suitable choice is given by
\beq
    \Phi_0 \,  = \, \prod_{1 \le i < j \le r} ( z_i  - z_j )^{ \frac{1}{\ka} \Om_{ij}}
    \label{6-24}
\eeq
which is defined on $\C_r$.
As studied in the previous sections,  elements of homology groups
satisfy the equivalent relation $d \log \Phi_0 \equiv 0$.
In particular, from (\ref{3-26}) we see that this enable us to replace
the ordinary derivative $\d_z$ by the covariant derivative $D_z$.
In a sense this equivalent relation can be thought of
as an origin of the minimal coupling principle in gauge theories.

In conclusion, the holonomy operator of the KZ connection
is essentially given by the integral $I_r$.
Motivated by the bilinear construction of the generalized
hypergeometric functions (which we have reviewed in the previous
sections), we analyze $I_r$ in terms of (non-twisted)
homology and cohomology group of the relevant physical configuration space
$\C_r = X_r / \S_r$ or its loop space $L \C_r$.
Following the notation in (\ref{1-18}), the construction can
be written as
\beqar
    H_r ( \C_r , {\bf R} ) \times H^r ( \C_r , \mathfrak{g}^{\otimes r} )
    & \longrightarrow & {\bf C}
    \label{6-25a} \\
    H_0 ( L \C_r , {\bf R} ) \times H^0 ( L \C_r , \mathfrak{g}^{\otimes r} )
    & \longrightarrow & {\bf C}
    \label{6-25b}
\eeqar
The two interpretations of the integral $I_r$ suggest that
an unambiguous constituent of $I_r$ is given by its integrand $I_r (z)$, {\it i.e.},
\beqar
    I_r (z) & = &
    \sum_{\si \in \S_{r-1} } \frac{1}{\ka^r}
    \frac{
     \Om_{1 \si_2} \Om_{\si_2 \si_3} \cdots \Om_{\si_r 1}
    }{
    ( z_1 - z_{\si_2} ) ( z_{\si_2} -  z_{\si_3} ) \cdots ( z_{\si_{r-1}} - z_{\si_r}  )  ( z_{\si_r} - z_1 )
    }
    \label{6-26a} \\
    I_r &=&
    \oint_\ga I_r (z) \, d z_1 \wedge d z_2 \wedge \cdots \wedge d z_r
    \label{6-26b} \\
    &=&
    \sum_{\si \in \S_{r-1}} \frac{1}{\ka^r} \oint_{\ga}  \Om_{1 \si_2} \Om_{\si_2 \si_3} \cdots \Om_{\si_r 1}
    \, \om_{1 \si_2} \wedge \om_{\si_2 \si_3} \wedge \cdots \wedge \om_{\si_r 1}
    \label{6-26c}
\eeqar
The expressions (\ref{6-26b}) and (\ref{6-26c}) (or (\ref{6-14})) correspond
to the interpretations (\ref{6-25b}) and (\ref{6-25a}), respectively.
Since the equation between (\ref{6-26b}) and (\ref{6-26c}) is mathematically subtle,
we need to further consider the meaning of this.
Obviously, the equation holds when the $r$-form
$\al_{\si}^{r} := d (z_1 - z_{\si_2} ) \wedge d ( z_{\si_2} - z_{\si_3} ) \wedge
\cdots \wedge d ( z_{\si_r} - z_1 )$ is equivalent to
$d z_1 \wedge \cdots \wedge d z_r$, regardless choices of $\si$.
Since $z_i$'s are coordinates of $\C_r$ they satisfy the permutation
invariance and $z_i - z_j \ne 0$ $(i \ne j)$. Thus natural non-vanishing
coordinates on $\C_r$ can be taken by
$ ( z_1 - z_{\si_2} ,  z_{\si_2} - z_{\si_3} , \cdots , z_{\si_r} - z_1 ) $.
These are the coordinates on $\C_r$, that is, these are supposed to
be independent of the permutation $\si$.
If we impose the condition $z_i \ne 0$, we can then identify
the above non-vanishing coordinates as $( z_1 , \cdots , z_r )$.
This means that $z_i$'s are the coordinates on $\cp^{r-1}$
rather than ${\bf C}^r$; remember that $\C_r = X_r / \S_r$ and
we have defined $X_r = {\bf C}^r   - \bigcup_{ i < j} {\cal H}_{ij}$ where
${\cal H}_{i j}  =  \{ (z_1 , \cdots , z_r )
\in {\bf C}^r ; \,  z_i - z_j = 0 \, (i \ne j)\}$.
This means that as far as we consider the projected complex
spaces we can equate the above $r$-form $\al_{\si}^{r}$ to
the simplest form $d z_1 \wedge \cdots \wedge d z_r$, and
the above relations (\ref{6-26b}) and (\ref{6-26c}) hold.

In terms of $I_r (z)$ {\it integrand} part of the holonomy operator $\Theta_\ga$
can be written as
\beq
    \Theta_\ga (z) \, = \,
    \exp \left( \sum_{ r \ge 2 } I_r (z)
    \right)
    \label{6-27}
\eeq

\vskip 0.5cm \noindent
\underline{Simplification of  the algebraic part}

Having considered analytic aspects of the holonomy operators, we now
consider simplification of the algebraic structure of $I_r$.
As emphasized earlier in this section, the algebra of the KZ connections
is determined by the infinitesimal braid relations
(\ref{3-6a}) and (\ref{3-6b}).
Using the $SL(2, {\bf C})$ algebra in (\ref{6-5}), we introduce
a bialgebraic operator \cite{Abe:2009kn}:
\beq
    A_{ij} = a_{i}^{(+)} \otimes a_{j}^{(0)} + a_{i}^{(-)} \otimes a_{j}^{(0)} \, .
    \label{6-28}
\eeq
$A_{ij}$ satisfy the infinitesimal braid relations (\ref{3-6a}), (\ref{3-6b}).
Thus, the operator
\beq
    A =  \frac{1}{\kappa} \sum_{1 \le i < j \le n} A_{ij} \, \om_{ij}
    \label{6-29}
\eeq
obeys the flatness condition
$DA = dA - A \wedge A = - A \wedge A = 0$
where $D$ is now a covariant exterior derivative $D = d - A$.
This relation guarantees the existence of
the holonomy operator for $A$.

The bialgebraic structures of  $\Om$ and $A$ are
different but the constituents of these remain the same, {\it i.e.}, they are given
by $a_{i}^{(0)}$ and $a_{i}^{(\pm)}$.
Thus, we can use the same Hilbert space $V^{\otimes n}$
and physical configuration $\C$ for both $\Om$ and $A$.
The KZ equation of $A$ is then given by
\beq
    D \Psi = (d - A) \Psi = 0
    \label{6-30}
\eeq
where $\Psi$ is a function of a set of complex variables $(z_1 , z_2 , \cdots z_n)$.
In analogy with (\ref{6-11}) the holonomy operator of $A$ can be defined as
\beq
    \Theta_{\ga}^{(A)} \, = \, \Tr_{\ga} \, \Path \exp \left[
    \sum_{r \ge 2} \oint_{\ga} \underbrace{A \wedge A \wedge \cdots \wedge A}_{r}
    \right]
    \label{6-31}
\eeq
The algebraic part of $\Theta_{\ga}^{(A)}$ can be simplified as follows.
We first note that the commutator $[A_{12}, A_{23}]$ can be calculated as
\beqar
    \nonumber
    [A_{12}, A_{23}]
    &=& a_{1}^{(+)} \otimes a_{2}^{(+)} \otimes a_{3}^{(0)}
    - a_{1}^{(+)} \otimes a_{2}^{(-)} \otimes a_{3}^{(0)}  \\
    && \!\!\! + \, a_{1}^{(-)} \otimes a_{2}^{(+)} \otimes a_{3}^{(0)}
    - a_{1}^{(-)} \otimes a_{2}^{(-)} \otimes a_{3}^{(0)}
    \label{6-32}
\eeqar
An analog of (\ref{6-13}) is then expressed as
\beqar
    &&
    \Path \oint_{\ga} \underbrace{A \wedge  \cdots \wedge A}_{r}
    \nonumber \\
    &=&
    \oint_{\ga}  A_{1 2} A_{2 3} \cdots A_{r 1}
    \, \om_{12} \wedge \om_{23} \wedge \cdots \wedge \om_{r 1} \\
    \nonumber
    &=& \frac{1}{2^{r}} \sum_{(h_1, h_2, \cdots , h_r)}
    (-1)^{h_1 + h_2 + \cdots + h_r}  \,
    a_{1}^{(h_1)} \otimes a_{2}^{(h_2)} \otimes \cdots \otimes a_{r}^{(h_r)}
    \, \oint_{\ga} \om_{12} \wedge \cdots \wedge \om_{r1}
    \nonumber \\
    \label{6-33}
\eeqar
where $h_{i}$ denotes $h_{i} = \pm = \pm 1$ ($i=1,2,\cdots, r$).
In the above expression, we define
$a_{1}^{(\pm)} \otimes a_{2}^{(h_2)} \otimes
\cdots \otimes a_{r}^{(h_r)} \otimes a_{1}^{(0)}$ as
\beqar
    \nonumber
    a_{1}^{(\pm)} \otimes a_{2}^{(h_2)} \otimes \cdots \otimes a_{r}^{(h_r)} \otimes a_{1}^{(0)}
    & := &
    \hf [a_{1}^{(0)} , a_{1}^{(\pm)}] \otimes a_{2}^{(h_2)} \otimes \cdots \otimes a_{r}^{(h_r)} \\
    &=&
    \pm \hf a_{1}^{(\pm)} \otimes a_{2}^{(h_2)} \otimes \cdots \otimes a_{r}^{(h_r)}
    \label{6-34}
\eeqar
The braid trace over (\ref{6-33}) is expressed as
\beqar
    \Tr_{\ga} \Path \oint_{\ga}
    \underbrace{A \wedge \cdots \wedge A}_{r}
    &=&
    \frac{1}{2^{r}}
    \sum_{(h_1, h_{2} , \cdots , h_{r} )}
    (-1)^{h_1 + h_{2} + \cdots + h_{r} }  \,
    a_{1}^{(h_1)} \otimes a_{2}^{(h_2)} \otimes \cdots \otimes a_{r}^{(h_r)}
    \nonumber \\
    && ~\times
    \sum_{\si \in \S_{r-1}}
    \oint_{\ga}  \om_{1 \si_2} \wedge \om_{\si_2 \si_3} \wedge \cdots \wedge \om_{\si_r 1}
    \label{6-35}
\eeqar
Applying the result (\ref{6-27}), we find that the {\it integrand} part
of the holonomy operator $\Theta_{\ga}^{(A)}$ is expressed as
\beq
    \Theta_{\ga}^{(A)} (z) \, = \,
    \exp \left( \sum_{ r \ge 2 } I_{r}^{(A)} (z)
    \right)
    \label{6-36}
\eeq
where
\beq
    I_{r}^{(A)} (z)  \, =
    \sum_{(h_1, h_{2} , \cdots , h_{r} )}
    \sum_{\si \in \S_{r-1}}
    \left( \frac{1}{2 \ka} \right)^r
    \frac{
    (-1)^{h_1 + h_{2} + \cdots + h_{r} }  \,
    a_{1}^{(h_1)} \otimes a_{2}^{(h_2)} \otimes \cdots \otimes a_{r}^{(h_r)}
    }{
    ( z_1 - z_{\si_2} ) ( z_{\si_2} -  z_{\si_3} ) \cdots ( z_{\si_{r-1}} - z_{\si_r}  )  ( z_{\si_r} - z_1 )
    }
    \label{6-37}
\eeq

\section{Holonomy formalism for gluon amplitudes}

In this section we first present an improved description of the
holonomy formalism \cite{Abe:2009kn}, using $\Theta_{\ga}^{(A)} (z) $
obtained in the previous section.
Essential ingredients of recent developments in the calculation
of gluon amplitudes are the spinor-helicity
formalism and the supertwistor space.
We begin with a brief review of these topics.

\noindent
\underline{Spinor-helicity formalism}

Spinor momenta of massless particles, such as gluons and
gravitons, are generally given by two-component complex spinors.
In terms of four-momentum $p_\mu$ $(\mu = 0,1,2,3)$, which
obey the on-shell condition $p^2 = \eta^{\mu\nu} p_\mu p_\nu =
p_{0}^{2} - p_{1}^{2} - p_{2}^{2} - p_{3}^{2} =0$,
$\eta^{\mu\nu}$ denoting the Minkowski metric,
the spinor momenta can be expressed as
\beq
u^A = \frac{1}{ \sqrt{p_0 - p_3}} \left(
        \begin{array}{c}
          {p_1 - i p_2} \\
          {p_0 - p_3} \\
        \end{array}
      \right) \, , ~~
\bu_\Ad  = \frac{1}{ \sqrt{p_0 - p_3}}
    \left(
         \begin{array}{c}
           {p_1 + i p_2 } \\
           {p_0 - p_3} \\
         \end{array}
    \right)
\label{7-1}
\eeq
where we follow a convention to express a spinor as a column vector.
The spinor momenta are two-component spinors; $A$ and $\Ad$ take values of $(1,2)$.
With these, the four-momentum $p_\mu$ can be parametrized as
a $(2 \times 2)$-matrix
\beq
    p^{A}_{\, \Ad}  = (\si^\mu)^{A}_{\, \Ad} \, p_\mu   =
        \left(
            \begin{array}{cc}
              p_0 + p_3 & p_1 - i p_2 \\
              p_1 + i p_2 & p_0 - p_3 \\
            \end{array}
         \right) =  u^A \bu_\Ad
\label{7-1a}
\eeq
where $\si^\mu = ( {\bf 1}, \si^i)$ where $\si^i$ ($i = 1,2,3$) denotes
the $(2 \times 2)$ Pauli matrices and
${\bf 1}$ is the $(2 \times 2)$ identity matrix.
Requiring that $p_\mu$ be real, we can take $\bu_\Ad$ as a
conjugate of $u^A$, {\it i.e.}, $\bu_\Ad = (u^A)^*$.

Lorentz transformations of $u^A$ are given by
\beq
    u^A \rightarrow (g u)^A
    \label{7-2}
\eeq
where $g \in SL(2, {\bf C})$
is a $(2 \times 2)$-matrix representation of $SL(2,{\bf C})$;
the complex conjugate of this relation leads to Lorentz transformations of $\bu_\Ad$.
Four-dimensional Lorentz transformations are realized by a
combination of these, that is, the four-dimensional Lorentz symmetry is
given by $SL(2,{\bf C}) \times SL(2,{\bf C})$.
Scalar products of $u^A$'s or $\bu_\Ad$'s, which are invariant under the
corresponding $SL(2,{\bf C})$, are expressed as
\beq
    u_i \cdot u_j \, := \, (u_i u_j) \, = \,  \ep_{AB} u_{i}^{A}u_{j}^{B} \, , ~~~~~
    \bu_i \cdot \bu_j \, := \, [\bu_i \bu_j]  \, = \, \ep^{\Ad \Bd} \bu_{i \, \Ad}
    \bu_{j \, \Bd}
    \label{7-3}
\eeq
where $\ep_{AB}$ is the rank-2 Levi-Civita tensor.
This can be used to raise or lower the indices, {\it e.g.}, $u_B = \ep_{AB}u^A$.
Notice that these products are zero when $i$ and $j$ are identical.

For a theory with conformal invariance, such as Maxwell's electromagnetic theory
and $\N = 4$ super Yang-Mills theory, we can impose scale invariance
on the spinor momentum, {\it i.e.},
\beq
    u^A \sim \la u^A \, , ~~~~~ \la \in {\bf C} - \{ 0 \}
    \label{7-4}
\eeq
where $\la$ is non-zero complex number.
With this identification, we can regard
the spinor momentum $u^A$ as a homogeneous coordinate
of the complex projective space $\cp^1$.
The local coordinate of $\cp^1$ is represented by
a complex variable $z \in {\bf C} - \{\infty \}$. This can be related to
$u^A$ by the following parametrization:
\beq
    u^A = \frac{1}{\sqrt{p_0 - p_3}} \left(
        \begin{array}{c}
          {p_1 - i p_2} \\
          {p_0 - p_3} \\
        \end{array}
      \right)
    := \left(
        \begin{array}{c}
          \al \\
          \bt \\
        \end{array}
      \right)
    = \al \left(
        \begin{array}{c}
          1 \\
          z \\
        \end{array}
      \right)
    ,~~~~ z = \frac{\bt}{\al} ~~ (\al \ne 0)
    \label{7-5}
\eeq
The local complex coordinate of
$\cp^1$ can be taken as $z = \bt / \al$ except in the vicinity of $\al = 0$,
where we can instead use $1/ z = \al / \bt$ as the local coordinate.

A helicity of a massless particle is generally determined by the
so-called Pauli-Lubanski spin vector. In the spinor-momenta formalism,
we can define an analog of this spin vector. This enables us to
define a helicity operator of massless particles as
\beq
    h = 1 -  \hf u^A \frac{\d}{\d u^A}
    \label{7-6}
\eeq
This means that the helicity of the particle
is determined by the degree of homogeneity in $u$.

\noindent
\underline{Twistor and supertwistor spaces}

Twistor space is defined by a four-component spinor
$W^I =( \pi^A, v_\Ad)$ $(I = 1,2,3,4)$ where $\pi^A$ and $v_\Ad$ are two-component
complex spinors.
From this definition, it is easily understood that
twistor space is represented by the complex homogeneous coordinates of
$\cp^3$. Thus, $W^I$ correspond to homogeneous coordinates of $\cp^3$ and satisfy
the following relation.
\beq
    W^I  \sim \la W^I \, , ~~~~~ \la \in {\bf C} - \{ 0 \}
    \label{7-7}
\eeq
In twistor space, the relation between $\pi^A$ and $v_\Ad$ is defined as
\beq
    v_\Ad = x_{\Ad A} \pi^A
    \label{7-8}
\eeq
where $x_{\Ad A}$ represent the local coordinates on $S^4$.
This can be understood from the fact that $\cp^3$ is a $\cp^1$-bundle over $S^4$.
We consider that the $S^4$ describes a four-dimensional compact spacetime.
Notice that in twistor space the spacetime coordinates $x_{\Ad A}$
are emergent quantities.
Four-dimensional diffeomorphisms, {\it i.e.},
general coordinate transformations, is therefore realized by
\beq
    u^A \rightarrow u'^A
    \label{7-8a}
\eeq
rather than $x_{\Ad A} \rightarrow {x'}_{\Ad A}$.

The key idea of the spinor-helicity formalism in twistor space
is the identification of the $\cp^1$ fibre of twistor space
with the $\cp^1$ on which the spinor momenta are defined.
In other words, we identify $\pi^A$ with the spinor momenta $u^A$
so that we can essentially describe four-dimensional physics
in terms of the coordinates of $\cp^1$.
In the spinor-momenta formalism,
the twistor-space condition $v_\Ad = x_{\Ad A} \pi^A$ is then expressed as
\beq
    v_\Ad \, = \, x_{\Ad A} u^A
    \label{7-9}
\eeq
In what follows we use the spinor momenta $u^A$ for the role of $\pi^A$
in twistor space.

Supertwistor space is defined by the homogeneous coordinates of $\cp^{3|4}$.
We can denote the coordinates by
\beq
    W^{\hat{I}} \, = \, ( u^A, v_\Ad, \xi^\al )
    \label{7-9a}
\eeq
where we introduce Grassmann variables
\beq
    \xi^\al \, = \, \th^{\al}_{A} u^A ~~~~~~ (\al = 1,2,3,4)
    \label{7-9b}
\eeq
in addition to the twistor variables $W^I =( u^A, v_\Ad)$ in (\ref{7-7}).
$I$ and $\widehat{I}$ are composite indices that can be
labeled as $I = 1,2,3,4$ and $\hat{I} = 1,2, \cdots , 8$, respectively.
Coordinates of a compact four-dimensional spacetime $x_{\Ad A}$ and
their chiral superpartners $\th^{\al}_{A}$ arise
from the supertwistor space with an imposition of the supertwistor conditions:
\beq
    v_\Ad = x_{A  \Ad} u^A \, , ~~~
    \xi^\al = \th_{A}^{\al} u^A
    \label{7-10}
\eeq
These are a supersymmetric analog of the twistor-space condition (\ref{7-9}).

As in the case of a superspace formalism, the coordinates $x_{\Ad A}$ can be
extended to $x_{\Ad A} \rightarrow x_{\Ad A} + 2 \bth_{\al \Ad} \th^{\al}_{A}$.
So a supersymmetric extension of the product $x_{\Ad A} p^{A \Ad}$ is expressed as
\beqar
    x_{\Ad A} p^{A \Ad} & \rightarrow &
    \left. x_{\Ad A} u^A \bu^\Ad +  2 \bth_{\al \Ad}  \th_{A}^{\al} u^A  \bu^\Ad
    \right|_{v_\Ad = x_{\Ad A} u^A , \, \xi^\al = \th^{\al}_{A} u^A } \nonumber \\
    &=& v_\Ad \bu^\Ad + 2 \bar{\et}_\al \xi^\al
    \label{7-10a}
\eeqar
where we use the supertwistor conditions (\ref{7-10}). We also define antiholomorphic
Grassmann variables $\bar{\et}_\al$ ($\al = 1, 2, 3, 4$) as
\beq
    \bar{\et}_\al = \bu_{\Ad}  \bth_{\al}^{\Ad}
    \label{7-10b}
\eeq

\vskip 0.5cm \noindent
\underline{Holonomy formalism for MHV amplitudes}

Having reviewed the spinor-helicity formalism in twistor space,
we now define creation operators of gluons.
As mentioned in (\ref{7-6}), the helicity of a particle is determined
by the degree of homogeneity in $u$.
In accordance with (\ref{7-6}), we define
the gluon operators $a_{i}^{( \pm )}$ of helicity $\pm$ and their superpartners as
\beqar
    a_{i}^{(+)} (\xi_i) &=& a_{i}^{(+)}
    \nonumber \\
    a_{i}^{\left( + \hf \right)} (\xi_i) &=& \xi_{i}^{\al}
    \, a_{i \al}^{ \left( + \hf \right)}
    \nonumber \\
    a_{i}^{(0)} (\xi_i) &=& \hf \xi_{i}^{\al} \xi_{i}^{\bt} \, a_{i \al \bt}^{(0)}
    \label{7-11} \\
    a_{i}^{\left(- \hf \right)} (\xi_i) &=& \frac{1}{3!} \xi_{i}^{\al}\xi_{i}^{\bt}\xi_{i}^{\ga}
    \ep_{\al \bt \ga \del} \, {a_{i}^{ \del}}^{ \left( - \hf \right)}
    \nonumber \\
    a_{i}^{(-)} (\xi_i) &=& \xi_{i}^{1} \xi_{i}^{2} \xi_{i}^{3} \xi_{i}^{4} \, a_{i}^{(-)}
    \nonumber
\eeqar
where $i=1,2,\cdots, n$ and $\hat{h}_{i} = (0 , \pm \hf , \pm )$
respectively denote the numbering index and the helicity of the particle.
The color factor of gluon can be attached to each of the physical operators:
\beq
    a_{i}^{(\hat{h}_{i})} \, = \, t^{c_i}   a_{i}^{(\hat{h}_{i}) c_i}
    \label{7-12}
\eeq
where $t^{c_i}$'s are given by the generators of the $SU(N)$ gauge group.

In the coordinate-space (or superspace) representation,
the physical operators can be expressed as
\beq
    a_{i}^{(\hat{h}_{i})} (x, \th)
    \, = \,
    \left. \int d\mu (p_i) ~ a_{i}^{(\hat{h}_{i})} (\xi_i) ~  e^{ i x_\mu p_{i}^{\mu} }
    \right|_{\xi_{i}^{\al} = \th_{A}^{\al} u_{i}^{A} }
    \label{7-13}
\eeq
where $d \mu (p)$ denotes the Nair measure:
\beq
    d \mu (p_i) \, = \, \frac{d^3 p_i}{ (2 \pi )^3 } \frac{1}{2 {p_{0}}_i }
    \, = \,
    \frac{1}{4} \left[
    \frac{u_i \cdot du_i }{2 \pi i} \frac{d^2 \bu_i}{(2 \pi)^2} -
    \frac{\bu_i \cdot d \bu_i}{2 \pi i} \frac{d^2 u_i}{(2 \pi)^2}
    \right]
    \label{7-14}
\eeq

The maximally helicity violating (MHV) amplitudes are the scattering amplitudes of $(n-2)$
positive-helicity gluons and $2$ negative-helicity gluons.
In a momentum-space representation, the MHV tree amplitudes
are expressed as
\beqar
    \A_{\rm MHV}^{(r_{-} s_{-})} (u, \bu)
    & = & i g^{n-2}
    \, (2 \pi)^4 \del^{(4)} \left( \sum_{i=1}^{n} p_i \right) \,
    \widehat{A}_{\rm MHV}^{(r_{-} s_{-})} (u)
    \label{7-15a}  \\
    \widehat{A}_{\rm MHV}^{(r_{-} s_{-})} (u) \! &=&
    \!\! \sum_{\si \in \S_{n-1}}
    \Tr (t^{c_1} t^{c_{\si_2}} t^{c_{\si_3}} \cdots t^{c_{\si_n}}) \,
    \frac{ (u_r u_s )^4}{ (u_1 u_{\si_2})(u_{\si_2} u_{\si_3})
    \cdots (u_{\si_n} u_1)}
    \label{7-15b}
\eeqar
where $r$ and $s$ denote
the numbering indices of the two negative-helicity gluons out of the total $n$ gluons.
$g$ is the coupling constant of gluon interactions.

Now it is straightforward to construct an S-matrix functional for
the MHV amplitudes by use of the integrand part of the
holonomy operator in (\ref{6-37}).
We first replace the operators $a_{i}^{(h_i ) }$ in (\ref{6-37}) by
$a_{i}^{(\hat{h}_{i})} (x, \th)$ in (\ref{7-13}).
We further use the spinor momenta $u_i$'s for the complex
variables $z_i$'s on $\cp^1$.
This leads to a supersymmetric versions of $I_{r}^{(A)} (z)$, {\it i.e.},
\beq
    I_{r}^{(A)}(u; x , \th ) \, = \,
    \sum_{( \hat{h}_1, \hat{h}_{2} , \cdots , \hat{h}_{r} )}
    \sum_{\si \in \S_{r-1}} g^r
    \frac{
    (-1)^{\hat{h}_1 + \hat{h}_{2} + \cdots + \hat{h}_{r} }  \,
    a_{1}^{(\hat{h}_1)}(x, \th)  \otimes \cdots \otimes a_{r}^{(\hat{h}_r)}(x, \th)
    }{
    ( u_1 u_{\si_2} ) ( u_{\si_2} u_{\si_3} ) \cdots ( u_{\si_{r-1}} u_{\si_r}  )  ( u_{\si_r} u_1 )
    }
    \label{7-16}
\eeq
where we define the coupling constant $g$ by
\beq
    g \, = \,  \frac{1}{2 \ka}
    \label{7-17}
\eeq
In terms of the supersymmetric version of $\Theta_{\ga}^{(A)} (z)$,
\beq
    \Theta_{\ga}^{(A)} (u; x , \th ) \, = \,
    \exp \left( \sum_{ r \ge 2 } I_{r}^{(A)} (u; x , \th )
    \right) ,
    \label{7-18}
\eeq
the S-matrix functional for the MHV tree amplitudes can be constructed as
\beq
    \F_{\rm MHV} \left[ a^{(\hat{h})c } \right]\, = \,
    \exp \left[
    \frac{i}{g^2} \int d^4 x d^8 \th \, \Theta_{\ga}^{(A)}(u; x , \th )
    \right]
    \label{7-19}
\eeq
Indeed we can check that $\F_{\rm MHV}$ generates the MHV amplitudes:
\beqar
    &&
    \left. \frac{\del}{\del a_{1}^{(+) c_1}} \otimes
    \cdots \otimes \frac{\del}{\del a_{r}^{(-) c_r}} \otimes
    \cdots \otimes \frac{\del}{\del a_{s}^{(-) c_s}} \otimes
    \cdots \otimes \frac{\del}{\del a_{n}^{(+) c_n}}
    ~ \F_{\rm MHV} \left[  a^{(h)c} \right] \right|_{a^{(h)c}=0}
    \nonumber \\
    &=&
    \prod_{i=1}^{n} \int d \mu (p_i ) \A_{\rm MHV}^{(r_{-} s_{-})} (u, \bu )
    \label{7-20}
\eeqar
where $a^{(h)c}$'s denote the gluon creation operators
in the momentum-space representation, which are treated as source
functions here.
In the above derivation we use the fact that
the Grassmann integral over $\th$'s vanishes unless we have
the following integrand:
\beq
    \left. \int d^8 \th  \, \xi_{r}^{1}\xi_{r}^{2}\xi_{r}^{3}\xi_{r}^{4}
    \, \xi_{s}^{1}\xi_{s}^{2}\xi_{s}^{3}\xi_{s}^{4}
    \right|_{\xi_{i}^{\al} = \th_{A}^{\al} u_{i}^{A} }
    = \, (u_r u_s )^4
    \label{7-21}
\eeq
This relation guarantees that only the MHV amplitudes are picked up
upon the execution of the Grassmann integral.

Notice that $\A_{\rm MHV}^{(r_{-} s_{-})} (u, \bu )$
contains the delta function
\beq
    (2 \pi )^4 \del^{(4)} ( p_1 + \cdots + p_n )
    \, = \, \int d^4 x ~ e^{i (p_1 + \cdots + p_n ) x }
    \label{7-22}
\eeq
Thus the resultant expression in (\ref{7-20})
can be proportional to the products of $\del^{(4)} (x)$'s.
Physically, this is obvious because $n$ gluons are supposed to
be scattering at a single point. But, mathematically, we need
to be careful about the treatment of the products of the delta
functions since there are no rigorous definitions for such products.
The problem may be solved if we consider in the momentum representation
where the holomorphic MHV amplitudes
$\widehat{A}_{\rm MHV}^{(r_{-} s_{-})} (u)$ in (\ref{7-15b}) can
be generated as
\beqar
    \nonumber
    &&
    \frac{\del}{\del a_{1}^{(+) c_1} (x_{1}) } \otimes
    \cdots \otimes \frac{\del}{\del a_{r}^{(-) c_r} (x_{r})} \otimes \cdots
    \\
    \nonumber
    && ~~~~~~~~~~~~~~~~
    \left. \cdots \otimes \frac{\del}{\del a_{s}^{(-) c_s} (x_{s})} \otimes
    \cdots \otimes \frac{\del}{\del a_{n}^{(+) c_n} (x_{n})}
    ~ \F_{\rm MHV} \left[ a^{(h)c} \right]
     \right|_{a^{(h)c} ( x ) =0}
    \\
    & = &
    i g^{n-2} \, \widehat{A}_{\rm MHV}^{(r_{-} r_{-})} (u)
    \label{7-23}
\eeqar
where $a^{(h)c} ( x )$'s play the same role as $a^{(h)c} $ in  (\ref{7-20})
except that they are now given by $x$-space representation of
the gluon creation operators:
\beq
    a_{i}^{(h_i)} (x) =
    \int d \mu (p_i)
    \,  a_{i}^{(h_i)} \,  e^{i x_\mu p_{i}^{\mu} }
    \label{7-24}
\eeq
As studied in \cite{Abe:2011af}, however, the expression
(\ref{7-20}) turns out to be more useful for the computation
of one-loop amplitudes.

\vskip 0.5cm \noindent
\underline{CSW rules and non-MHV amplitudes in holonomy formalism}

General amplitudes, the so-called non-MHV amplitudes, can be expressed in terms of
the MHV amplitudes $\widehat{A}_{MHV}^{(r_{-} s_{-})} (u)$ at tree level.
Prescription for these expressions is called the Cachazo-Svrcek-Witten (CSW) rules
\cite{Cachazo:2004kj}.
For the next-to-MHV (NMHV) amplitudes, which contain three negative-helicity
gluons, the CSW rules can be expressed as
\beq
    \widehat{A}^{(r_- s_- t_-)}_{\rm NMHV} (u) = \sum_{(i,j)}
    \widehat{A}^{(i_+ \cdots r_- \cdots s_- \cdots j_+ k_+)}_{\rm MHV} (u)
    \, \frac{\del_{kl}}{q_{ij}^2} \,
    \widehat{A}^{(l_- \, (j+1)_+ \cdots t_- \cdots (i-1)_+)}_{\rm MHV} (u)
    \label{7-25}
\eeq
where the sum is taken over all possible choices for $(i, j)$ that
satisfy the ordering $i < r < s < j < t$. The momentum transfer
$q_{ij}$ between the two MHV vertices is given by
\beq
    q_{ij} = p_i + p_{i+1 } + \cdots + p_{r} + \cdots + p_{s} + \cdots + p_{j}
    \label{7-26}
\eeq
where $p$'s denote four-momenta of gluons as before.
General non-MHV amplitudes are then obtained by an iterative use of
the relation (\ref{7-25}).

In terms of the expression (\ref{7-20})
the CSW rules can be implemented by a contraction operator
\beqar
    \widehat{W}^{(A)} (x) & = & \exp \left[-
    \int d \mu (q) \left(
    \frac{\del}{\del a_{p}^{(+)}} \otimes \frac{\del}{\del a_{-p}^{(-)}}
    \right)
    e^{- iq (x -y) }
    \right]_{y \rightarrow x}
    \nonumber  \\
    &=&
    \exp \left[-
    \int \frac{ d^4 q}{(2 \pi)^4} \frac{i}{q^2} \left(
    \frac{\del}{\del a_{p}^{(+)}} \otimes \frac{\del}{\del a_{-p}^{(-)}}
    \right)
    e^{- iq (x -y) }
    \right]_{y \rightarrow x}
    \label{7-27}
\eeqar
where we take the limit $y \rightarrow x$ with $x^0 - y^0 \rightarrow 0_+$
(keeping the time ordering $x^0 > y^0$).
$q$ denotes a momentum transfer of a virtual gluon.
As explicitly parametrized in (\ref{7-26}), this is off-shell quantity $q^2 \ne 0$.
Its  on-shell partner can be defined as
\beq
    q_\mu \, = \, p_\mu + w \eta_\mu
    \label{7-28}
\eeq
where $\eta_\mu$ is a reference null vector ($\eta^2 = 0$) and $w$ is a real parameter.
In (\ref{2-27}) $a_{p}^{(\pm)}$ denote the creation operators of a pair of virtual
gluons at ends of a propagator or at MHV vertices.
In the calculation of (\ref{2-27}) we also use the well-known identity
\beq
    \int d \mu ( q ) \left[
    \th ( x^0 - y^0 ) e^{-i q (x - y) } + \th ( y^0 - x^0 ) e^{i q (x - y)}
    \right]
    \, = \,
    \int \frac{ d^4 q} {(2 \pi )^4} \,
    \frac{i}{ q^2 + i \ep }
    \, e^{- i q( x - y ) }
    \label{7-29}
\eeq
where $\ep$ is a positive infinitesimal.

Using the above contraction operator, we can define an S-matrix functional
for general non-MHV amplitudes as
\beq
    \F \left[  a^{(h)c}  \right]
    \, = \,  W^{(A)} (x) \F_{\rm MHV} \left[  a^{(h)c} \right]
    \label{7-30}
\eeq
Generalization of the expression (\ref{7-20}) can be written as
\beqar
    &&
    \left. \frac{\del}{\del a_{1}^{(h_1) c_1}  } \otimes
    \frac{\del}{\del a_{2}^{(h_2) c_2} } \otimes
    \cdots \otimes \frac{\del}{\del a_{n}^{(h_n) c_n} }
    \, \F \left[  a^{(h)c}  \right] \right|_{a^{(h)c} =0}
    \nonumber \\
    & = &
    \prod_{i=1}^{n} \int d \mu (p_i) ~
    \A^{(1_{h_1} 2_{h_2} \cdots n_{h_n})}_{\rm N^{k}MHV} (u, \bu )
    \label{7-31}
\eeqar
where $\A^{(1_{h_1} 2_{h_2} \cdots n_{h_n})}_{\rm N^{k}MHV} (u, \bu )$
denotes a non-MHV version of the gluon amplitudes in the form of (\ref{7-15a}).
This is called N$^{k-2}$MHV amplitudes
where $h_i = \pm$ denotes the helicity of the $i$-th gluon, with
the total number of negative helicities being $k$. ($k = 2, 3, \cdots , n-2$)
Note that the expression (\ref{7-31}) is not necessarily limited to tree-level
amplitudes; for details on applications of the CSW rules to one-loop amplitudes
in the holonomy formalism, see \cite{Abe:2011af}.
The above formulation illustrates that  the holonomy operator of the KZ connection
plays an essential role in the construction of  the S-matrix functional
$\F \left[  a^{(h)c}  \right]$ for gluon amplitudes at least at tree level.
One of the main purposes for the present note is to study mathematical
foundations of the holonomy operator (which we have carried out in
the previous section) so as to give an improved description of the holonomy
formalism.

\section{Grassmannian formulations of gluon amplitudes}

In the following we briefly review more {\it powerful} formulations
of gluon amplitudes, known as Grassmannian formulations
\cite{oai:arXiv.org:0907.5418}-\cite{oai:arXiv.org:1006.1899}.
These are powerful, in particular, in computation of higher loop amplitudes
but we here consider basic tree-level formulations.

\vskip 0.5cm \noindent
\underline{The supertwistor conditions as the momentum conservation laws}

We first review various representations of gluon amplitudes, following \cite{Abe:2009kn}.
The twistor space condition $v_\Ad  =  x_{\Ad A} u^A$ in (\ref{7-9})
naturally leads to the relations
\beq
    v_\Ad \bu^\Ad \, = \, x_{\Ad A} p^{A \Ad}
    \, = \, 2 (x_0 p_0 - x_1 p_1 - x_2 p_2 - x_3 p_3 ) = 2 x_{\mu} p^{\mu}
    \label{8-1}
\eeq
where we use the rules of scalar products
(\ref{7-3}) between the spinor momenta. We also parametrize
$x_{\Ad A}$ in terms of the Minkowski coordinates as
\beq
    x_{\Ad A} = x_\mu (\si^\mu)_{A \, \Ad}  =
        \left(
            \begin{array}{cc}
              x_0 + x_3 & x_1 - i x_2 \\
              x_1 + i x_2 & x_0 - x_3 \\
            \end{array}
         \right)
    \label{8-2}
\eeq
The product (\ref{8-1}) suggests that the physical
phase space can be spanned by $( v_\Ad , \bu^\Ad )$,
rather than $( x_\mu , p^\mu )$, in twistor space.
We can then relate a function of $(u, \bu)$ to
a function of $(u, v)$ by Fourier transform integrals
\beq
    f(u, v) = \frac{1}{4}
    \int \frac{d^2 \bu}{(2 \pi )^2} \, f(u, \bu) \, e^{\frac{i}{2} v_\Ad \bu^\Ad}
    \, , ~~~~~
    f(u, \bu) =  \int d^2 v \, f( u, v) \,  e^{-\frac{i}{2} v_\Ad \bu^\Ad}
    \label{8-3}
\eeq
Similarly, by taking conjugates of these, we have
\beq
    f( \bar{v}, \bu ) = \frac{1}{4}
    \int \frac{d^2 u}{(2 \pi )^2} \, f(u, \bu) \, e^{\frac{i}{2} {\bar v}_A u^A }
    \, , ~~~~~
    f(u, \bu) = \int d^2 \bar{v} \, f( \bar{v}, \bu )\, e^{-\frac{i}{2} {\bar v}_A u^A }
    \label{8-4}
\eeq
These integrals are referred to as Fourier transforms in twistor space.

What is remarkable about the use of supertwistor space in gluon amplitudes
is that the supertwistor conditions (\ref{7-10}) automatically arise from
the momentum conservation. We shall review this point in the following.
We start from $n$-point N$^{k-2}$MHV amplitudes of the form
\beq
    \A_{n,k} (u, \bu)
    \, = \, i g^{n-2}
    \, (2 \pi)^4 \del^{(4)} \left( \sum_{i=1}^{n} p_i \right) \,
    \widehat{A}_{n,k} (u , \bu)
    \label{8-5}
\eeq
This is a generalized version of $\A_{\rm MHV}^{(r_{-} s_{-})} (u, \bu)$
in (\ref{7-15a}). Notice that $\widehat{A}_{n,k} (u , \bu)$ is no more
holomorphic to $u^A$'s for $k \ge 1$.
The momentum conservation is realized by the delta function:
\beq
    (2 \pi)^4 \del^{(4)} \left( \sum_{i=1}^{n} p_i \right)
    \, = \,
    \int d^4 x ~ e^{- i \, x_\mu \sum_i p_{i}^{\mu}}
    \, = \, \int d^4 x ~ e^{- \frac{i}{2} \, x_{\Ad A} \sum_i u_{i}^{A} \bu_{i}^{\Ad}}
    \label{8-6}
\eeq
We now introduce a fermionic partner of the momentum conservation:
\beq
    \del^{(8)} \left( \sum_{i=1}^{n} p_{i \, \Ad}^{A} \bth_{\al}^{\Ad} \right)
    \, = \, \del^{(8)} \left( \sum_{i=1}^{n} u_{i}^{A} \bar{\et}_{i \al}  \right)
    \, = \, \int d^8 \th ~ e^{- i \, \th_{A}^{\al}
    \sum_i u_{i}^{A} \bar{\et}_{i \al}}
    \label{8-7}
\eeq
where $\bar{\et}_\al = \bu_{\Ad}  \bth_{\al}^{\Ad}$ as defined in (\ref{7-10b}).
Adding this fermionic delta function, the amplitudes can be
represented by $(u, \bu ,\bar{\et} )$:
\beq
    \A_{n,k} (u, \bu, \bar{\et})
    \, = \,
    \del^{(8)} \left( \sum_i u_{i}^{A} \bar{\et}_{i \al} \right)
    \, \A_{n,k} (u, \bu)
    \label{8-8}
\eeq
The amplitudes can of course be expressed in terms of the supertwistor variables
$W^{\hat{I}}  =  ( u^A, v_\Ad, \xi^\al )$ in (\ref{7-9a}).
Using Fourier transforms in supertwistor space, we can obtain such representations as
\beqar
    \A_{n,k} (u, v, \xi)
    &=&
    \left[ \prod_{i=1}^{n}  \frac{1}{4} \int
    \frac{d^2 \bu_i}{(2 \pi)^2} d^4 \bar{\et}_i \right]
    \A_{n,k} (u, \bu, \bar{\et}) \, \exp \left(
    \frac{i}{2} \sum_{i} v_{i \Ad} \bu_{i}^{\Ad}
    + i \sum_{i} \bar{\et}_{i \al} \xi_{i}^{\al}
    \right)
    \nonumber \\
    &=&
    i g^{n-2}
    \int d^4 x d^8 \th \,
    \prod_{i=1}^{n} \del^{(2)} ( v_{i \Ad} - x_{\Ad A} u_{i}^{A} ) \,
    \del^{(4)} (\xi_{i}^{\al} - \th_{A}^{\al} u_{i}^{A} ) ~
    \widehat{A}_{n,k} (u, \bu )
    \label{8-9}
\eeqar
where we use (\ref{8-6}) and (\ref{8-7}).
The inverse transformation is given by
\beq
    \A_{n,k} (u, \bu, \bar{\et}) \, = \,
    \left[
    \prod_{i=1}^{n}
    \int d^2 v_i d^4 \xi_i
    \, e^{-\frac{i}{2} v_{i \Ad} \bu_{i}^{\Ad}} \,
    e^{- i \bar{\et}_{i \al} \xi_{i}^{\al} }
    \right] ~ \A_{n,k} (u,v, \xi)
    \label{8-10}
\eeq

Notice that the supertwistor conditions (\ref{7-10}) are indeed embedded
in $\A_{n,k} (u,v, \xi)$.
This illustrates intimate connections between supertwistor space and
massless gauge bosons.
In the holonomy formalism we consider the physical operators
in the coordinate-space representation, see (\ref{7-13}) or (\ref{7-24}),
and put the supertwistor conditions by hand. This formulation is suitable
as far as we rely on the CSW rules where the amplitudes factorize
into the MHV vertices $\widehat{A}_{m,1} (u, \bu )$ ($m \le n$)
which are holomorphic to $u^A$'s, apart from
the contributions from the momentum conservation.
To describe the non-MHV amplitudes in a more democratic manner,
we need to handle the non-holomorphic part of the amplitudes
properly. This can not be done in the holonomy formalism.
The most promising formulation is given by the
Grassmannian formulations.

\vskip 0.5cm \noindent
\underline{Grassmannian formulations of gluon amplitudes}

In the Grassmannian formulations the gluon amplitudes are
originally considered in terms of the so-called {\it dual}
supertwistor variables
\beq
    {\cal W}^{\hat{I}}
    \, = \, ( \bar{v}_A , \bu^\Ad , \bar{\et}_\al )
    \label{8-11}
\eeq
The relevant amplitudes $\A_{n,k} ( \bar{v} , \bu , \bar{\et} )$
can be Fourier transformed into
$\A_{n,k} (u, \bu, \bar{\et})$ in (\ref{8-10}) as
\beq
    \A_{n,k} (u, \bu, \bar{\et}) \, = \,
    \left[
    \prod_{j=1}^{n}
    \int d^2 \bar{v}_j
    \, e^{-\frac{i}{2} \bar{v}_{jA} u_j^A  }
    \right] ~ \A_{n,k} ( \bar{v} , \bu , \bar{\et} )
    \label{8-12}
\eeq
Stripping the fermionic part, we can rewrite (\ref{8-12}) as
\beqar
    \A_{n,k} (u, \bu, \bar{\et})
    &=&
    i g^{n-2}
    \, (2 \pi)^4 \del^{(4)} \left( \sum_{j=1}^{n} u_j^A \bu^{j \Ad} \right) \,
    \del^{(8)} \left( \sum_{j=1}^{n} u_{j}^{A} \bar{\et}_{j \al} \right)
    \widehat{A}_{n,k} (u , \bu)
    \label{8-13a} \\
    \widehat{A}_{n,k} ( u ,\bu )
    & = &
    \left[
    \prod_{j=1}^{n}
    \int d^2 \bar{v}_j
    \, e^{-\frac{i}{2} \bar{v}_{jA} u_j^A  }
    \right]    ~ \widehat{A}_{n,k} ( \bar{v} , \bu )
    \label{8-13b}
\eeqar
One of the key ideas of the Grassmannian formulation is that
the $\bar{v}$-dependence of
$\widehat{A}_{n,k} ( \bar{v} , \bu ) $
is factored out by a delta function as follows:
\beq
    \widehat{A}_{n,k} ( \bar{v} , \bu )
    \, = \, \delta \left( \sum_{j = 1}^{n} \sum_{i=1}^{k} z_{ij} \, \rho_i^A \, \bar{v}_{jA}
    \right) ~ \widehat{A}_{n,k} ( \bu )
    \label{8-14}
\eeq
where $z_{ij}$ ($i = 1, 2, \cdots , k \, ; \, j = 1,2,\cdots , n \, ;
\, k = 2, 3, \cdots , n-2 $)
denotes an element of $( k \times n)$ complex matrix $Z$.
The delta function is analogous to the momentum-conservation delta functions
but it is qualitatively different from them as it involves mixing
of the numbering indices.
In (\ref{8-14}) $\rho_i^A$'s are considered as another set of the holomorphic
spinors (which are not necessarily spinor momenta).
Performing the $\bar{v}$-integral, $\widehat{A}_{n,k} ( u ,\bu ) $ are then expressed as
\beq
    \widehat{A}_{n,k} ( u ,\bu )
    \, = \,
    \prod_{j=1}^{n} \del^{(2)}
    \left( \sum_{i=1}^{k}  z_{ij} \, \rho_i^A  -  u_j^A \right)
    ~ \widehat{A}_{n,k} ( \bu )
    \label{8-15}
\eeq
Together with (\ref{8-13a}), this means that on top of the supertwistor conditions
we further impose the relation
\beq
    u_j^A \, = \, z_{1j} \, \rho_1^A  \, + \,  z_{2j} \, \rho_2^A  \, +  \,
    \cdots  \, +  \, z_{kj}  \, \rho_k^A
    ~~~~ (j= 1,2, \cdots ,n )
    \label{8-16}
\eeq

As mentioned in (\ref{7-8a}), the four-dimensional spacetime
is an emergent concept and the spinor momenta are more
fundamental quantities in twistor space.
In a multigluon system there is a possibility to extend our perspective
on the spinor momenta;
we may express one entry of the spinor momenta in terms
of another set of spinor momenta which are
given by two-component holomorphic spinors.
The extra condition (\ref{8-16}) exactly realizes this
possibility. One of the remarkable features in the
Grassmannian formalism is that the number of
linearly independent $\rho_i^A$'s is related
to the number of negative helicity gluons
in the amplitudes.
This makes it possible to describe non-MHV amplitudes
in a universal fashion.

Integrating over $\rho_i^A$'s, we see that the above expressions lead
to the Grassmannian formulations of the gluon amplitudes.
Apart from the color factor, the $n$-point N$^{k-2}$MHV gluon amplitudes
in the Grassmannian formulation are conjectured in a form of \cite{oai:arXiv.org:0907.5418}:
\beq
    {\cal L}_{n,k} ( \W ) \, = \,
    \int \frac{ d^{k \times n} Z}{ {\rm vol} [GL (k) ] }
    \frac{ \prod\limits_{i = 1}^{k} ~ \del^{4 | 4 }
    \left( \sum\limits_{j=1}^{n} \, z_{ij} \, \W_{j}^{\hat{I}} \right)}
    { (12 \cdots k)(2 3 \cdots  k+1) \cdots (n \, n+1 \, \cdots n+k-1 ) }
    \label{8-17}
\eeq
where $(j_1 j_2 \cdots j_k)$'s denote $k$-dimensional minor determinants
of $Z$ which consist of the $j_1$-th to $j_k$-th columns.
As explicitly shown in (\ref{8-17}), $j$'s are ordered such that
$(j_1 j_2 \cdots j_k)$ are cyclic to $(12 \cdots k)$ in mod $n$.
Thus the above minor determinant is
an {\it ordered} $k$-dimensional minor determinant of the
$k \times n$ matrix $Z$, which can be computed as
\beq
    ( j_1 j_2 \cdots j_k ) \,  := \, \sum_{\si \in \S_k}
    \left( \!\!
    \begin{array}{c}
      1 \, ~ 2 ~ \cdots ~ k \\
      \si_1 \si_2 \cdots \si_k \\
    \end{array} \!\!
    \right)
    \, z_{ \si_1 j_1} \, z_{ \si_2 j_2} \, \cdots \,
    z_{ \si_k j_k}
    \label{8-19}
\eeq
This is nothing but a Pl\"{u}cker coordinate of
the Grassmannian space $Gr ( k, n )$.
The delta function in (\ref{8-17}) is defined as
\beq
    \del^{4 | 4 } \left( \sum_{j=1}^{n} z_{ij} \, \W_{j}^{\hat{I}} \right)
    \, = \,
    \del^{(2)} \! \left( \sum_{j=1}^{n} z_{ij} \, \bar{v}_{jA} \right) \,
    \del^{(2)} \! \left( \sum_{j=1}^{n} z_{ij} \, \bu_{j}^{A} \right) \,
    \del^{(4)} \! \left( \sum_{j=1}^{n} z_{ij} \, \bar{\et}_{j \al} \right)
    \label{8-20}
\eeq
In the Grassmannian formulation (\ref{8-17}) the physical configuration
is given by $Z /GL ( k , {\bf C})$.
As discussed in (\ref{1-33}) (see also (\ref{1-20})), this
is equivalent to the Grassmannian space $Gr ( k , n)$.
Following the lines of arguments on $Gr ( k, n)$ in Section 2, it
is natural to consider $Gr ( k , n )$ as a configuration space
of $n$ hyperplanes in $\cp^{k-1}$. The modulo of $GL( k , {\bf C})$
is essentially required by avoiding the redundancy in
the configuration of $n$ hyperplanes. This redundancy is sometimes called
``gauge'' symmetry in the Grassmannian formulation.
With the redundancy eliminated, we also find that
the Grassmannian space $Gr ( k ,n )$ is represented by
$n$ distinct points in $\cp^{k-1}$.
The integral measure $d^{k \times n} Z /  {\rm vol} [GL (k) ] $ is relevant
to this degrees of freedom.

\vskip 0.5cm \noindent
\underline{Homology and cohomology interpretations of ${\cal L}_{n,k} ( \W )$}

The physical configuration space
of ${\cal L}_{n,k} ( \W )$ is analogous to that of
of the KZ solutions in (\ref{3-3}) and (\ref{3-4})
but in the present case the number of variables extends
from $n$ to $k \times n$ and the hyperplanes are defined by
$(j_1 j_2 \cdots j_k) = 0$. The physical configuration space in
the Grassmannian formulation is then expressed as
\beq
    X_{k \times n} \, = \, {\bf C}^{k \times n}  -
    \bigcup_{j = 1}^{n} {\cal H}_{(j \, j+1 \cdots j+k-1 )}
    \label{8-21}
\eeq
where, using the notation (\ref{8-19}),
${\cal H}_{(j \, j+1 \cdots j+n-1 )}$ is defined as
\beq
    {\cal H}_{(j \, j+1 \cdots j+k-1 )} \, = \,
     \{ Z \in Mat_{k , n} ( {\bf C} ) \, | ~  (j \, j+1 \cdots j+k-1 ) = 0 \}
    \label{8-22}
\eeq

Notice that there are no extra parameters except $z_{ij}$'s.
Following the arguments in Section 6 (see (\ref{6-18}) and below),
the cohomology group of interests in the present case
is the 0-dimensional cohomology group of a loop space $L \C_{k \times n}$ in $\C_{k \times n}$.
By definition an element of the 0-dimensional cohomology group of $L \C_{k \times n}$
can be expressed as
\beqar
    \varphi_{1,2, \cdots , n }
    &=&
    d \log (12 \cdots k ) \wedge d \log (2 3 \cdots k +1 ) \wedge \cdots \wedge
    d \log ( n 1 \cdots k-1)
    \nonumber \\
    &=&
    \frac{1}{(12 \cdots k )(2 3 \cdots k +1 ) \cdots ( n 1 \cdots k-1) }
    \frac{ d^{k \times n} Z}{ {\rm vol} [GL (k) ] }
    \label{8-23}
\eeqar
where we use the fact that $(12 \cdots k )$'s are the Pl\"{u}cker coordinates of $Gr ( k, n)$.
Thus, apart from the delta functions (\ref{8-20}), the integrand of
${\cal L}_{n,k} ( \W )$ is given by the element of $H^0 ( L \C_{k \times n} , {\bf R})$.
As discussed in Section 6, it is more natural to consider (\ref{8-23}) as
an element of $H^{k \times n} ( \C_{k \times n} , {\bf R})$, the $(k \times n )$-dimensional
cohomology group of $\C_{k \times n}$.
In either case, the dual supertwistor variables enter in a form of delta functions.
Thus, to be more precise, the integrand of ${\cal L}_{n,k} ( \W )$ is
given by the element of either $H^0 ( L \C_{k \times n} , \W)$ or
$H^{k \times n} ( \C_{k \times n} , \W )$.

Correspondingly, we can naturally interpret the contour of ${\cal L}_{n,k} ( \W )$ as
an element of $H_0 ( L \C_{k \times n} , {\bf R})$ or
$H_{k \times n} ( \C_{k \times n} , {\bf R} )$.
As considered in (\ref{6-17}), an element of
$H_{k \times n} ( \C_{k \times n} , {\bf R} )$
can be considered as a path in $\C_{k \times n}$ connecting
$n$ hyperplanes $H_{(j \, j+1 \, \cdots \, j+k-1)}$.
This is equivalent to say that the element is given by
that of the braid group $\B_n$:
\beq
    \ga \, \in \, H_{k \times n} ( \C_{k \times n} , {\bf R} )
    \, \cong \, \B_{n}
    \label{8-24}
\eeq
This is also in accord with the relation $\Pi_1 ( \C_{k \times n} )
\cong H_0 ( L \C_{k \times n} )$ in (\ref{6-18}). Namely,
we also have
\beq
    \ga \, \in \, H_0 ( L \C_{k \times n} , {\bf R}) \, \cong \,
    \Pi_1 ( \C_{k \times n} )\, \cong \, \B_{n}
    \label{8-25}
\eeq

In summary, following the notation in (\ref{1-18}),
we can also construct the integral ${\cal L}_{n,k} ( \W )$ as bilinear forms
\beqar
    H_{k \times n} ( \C_{k \times n} , {\bf R} ) \times
    H^{k \times n} ( \C_{k \times n} , \W )
    & \longrightarrow & {\bf C}
    \label{8-26a} \\
    H_0 ( L \C_{k \times n} , {\bf R} ) \times
    H^0 ( L \C_{k \times n} , \W )
    & \longrightarrow & {\bf C}
    \label{8-26b}
\eeqar

\section{Conclusion}

Recent developments in the computation of gluon amplitudes
provide an intriguing platform for interrelations between
modern physics and mathematics.
For example, the Grassmannian formulations of gluon amplitudes
shed new light on the notion of spacetime and suggest
the importance of Grassmannian spaces for our understanding
of gluons or particles themselves.
Recently, along the lines of these developments, interests
in generalized hypergeometric functions on the Grassmannian
spaces have been revived.
Naively, one would think that a suitable description of a multi-particle
system may be given by analytic functions of several complex variables.
The generalized hypergeometric functions provide a useful tool
to deal with such functions in a form of integrals, which can be constructed by use of
the concepts of twisted homology and cohomology.
One of the main purposes of the present note is to familiarize ourselves
to these concepts and apply them to physical formulations of gluon amplitudes.

Since the generalized hypergeometric functions are not
well recognized among physicists, we make this note sort of pedagogic.
We first review the definition of Aomoto's generalized
hypergeometric functions on $Gr(k+1, n+1)$, interpreting
their integral representations
in terms of twisted homology and cohomology.
We then consider reduction of the general $Gr(k+1 , n+1)$ case to
particular $Gr(2, n+1)$ cases. The case of $Gr( 2, 4)$ leads to
Gauss' hypergeometric functions. We carry out a thorough study
of this case in section 4.
Much of the present note, by nature, deals with reviews of
existed literature. But the results in (\ref{2-57})-(\ref{2-59})
are new as far as the author notices.

The case of $Gr( 2, 4)$ corresponds to a four-point solution
of the Knizhnik-Zamolodchikov (KZ) equation. The $Gr(2, n+1)$ cases
in general lead to $(n+1)$-point solutions of the KZ equation.
In section 5 we have reviewed these solutions. We further find
some ambiguities to relate the cases of $Gr ( k+1 , n+1 )$ to
the previously known Schechtman-Varchenko integral representations of
the KZ equation.
A system defined by the KZ equations provides a useful
description for a multi-particle systems. Since the monodromy representation
of the KZ equation is given by the braid group, the KZ system
is advantageous especially to analyze the multi-particle
system by use of braid groups.
The monodromy representation is also given by the
holonomy of the KZ connection, which can be expressed
in terms of the iterated integral.
In section 6 we review the definition of the holonomy operator
of the KZ connection.
We find that the integral representation of the
holonomy operator can be constructed
by a set of homology and cohomology groups in two
different ways. This interpretation is schematically summarized
in (\ref{6-25a}) and (\ref{6-25b}).

Equipped with this mathematical construction of the
holonomy operator, in section 7, we present an improved
review of the holonomy formalism for gluon amplitudes.
We also carry out a similar analysis on the Grassmannian formulations
of gluon amplitudes in section 8.
After a detailed introduction to the dual supertwistor variables ${\cal W}^{\hat{I}}$,
we review the definition of the integral representation of the amplitudes
${\cal L}_{n,k} ( \W )$ in (\ref{8-17}).
We find that the integral can also be constructed
in bilinear forms in terms of homology and cohomology group of
the relevant physical configuration space; see (\ref{8-26a}) and (\ref{8-26b}).
From these analyses we observe that the integral contours of ${\cal L}_{n,k} ( \W )$
are given by elements of $\B_n$, the braid group on $n$ strands.


\end{document}